\documentclass[11pt,a4paper]{article}
\usepackage{amsfonts}
\usepackage{amsmath,amssymb}
\usepackage{epsfig,graphicx}

\begin{document}

\begin{center}
{\LARGE \bigskip }{\Large SU(8) Grand Unification from Composite Quarks and
Leptons}

{\huge \bigskip }

\bigskip \bigskip

\textbf{\ J.L.~Chkareuli}$^{a,b}$

\bigskip

$^{a}$\textit{Institute of Theoretical Physics, Ilia State University, 0162
Tbilisi, Georgia \\[0pt]
}$^{b}$\textit{Andronikashvili Institute of Physics, Tbilisi State
University, 0186 Tbilisi, Georgia\ }

\bigskip \bigskip \bigskip \bigskip \bigskip \bigskip \bigskip

\textbf{Abstract}

\bigskip
\end{center}

Despite many interesting attempts in the past, both theoretical and
experimental, a possible composite structure of quarks and leptons is still
an open problem. Meanwhile, some observed replications in their spectra,
such as an existence of several quark color triplets, as well as the quark
and lepton weak doublets, and as many as three identical quark-lepton
families could mean that that there might exist the truly elementary
fermions, preons, being actual carriers of all the fundamental quantum
numbers involved and composing the observed quarks and leptons at larger
distances. Generally, certain regularities in replications of particles may
signal about their constituent structure. Indeed, just regularities in the
spectroscopy of hadrons observed in the nineteen-sixties made it possible to
discover the quark structure of hadrons in the framework of the so-called
Eightfold Way. We show that an inspirational Eightfold Way idea looks much
more relevant when applied to a next level of matter elementarity, namely,
to elementary preons and composite quarks and leptons. As we find below,
just the eight preon model and their local metaflavor symmetry $SU(8)_{MF}$
may under certain natural conditions determine the fundamental entities of
physical world at small distances and its total internal symmetry. Being
exact \ for preons, it gets broken for composites down to a conventional $%
SU(5)$ GUT with an extra local family symmetry $SU(3)_{F}$ and three
standard families of quarks and leptons. Though a tiny confinement scale for
universal preons composing both quarks and leptons makes it impossible to
directly confirm their composite nature, a simultaneous emergence of several
extra $SU(5)\times SU(3)_{F}$ multiplets of heavy composite fermions may
help with a model verification.

\bigskip


\bigskip

\thispagestyle{empty}\newpage

{\LARGE Contents}{\Huge \ \bigskip \bigskip \bigskip \bigskip }

\textbf{1 \ \ \ Introduction} \ \ \ \ \ \ \ \ \ \ \ \ \ \ \ \ \ \ \ \ \ \ \
\ \ \ \ \ \ \ \ \ \ \ \ \ \ \ \ \ \ \ \ \ \ \ \ \ \ \ \ \ \ \ \ \ \ \ \ \ \
\ \ \ \ \ \ \ \ \ \ \ \ \ \ \ \ \ \ \ \ \ \ \ \ \ \ \ \ \ \ \ \ \ \ \ \ \ \
\ \ \ \ \ \ \ \ \ \ \ \ \ \ \ \ \ \ \ \ \ \ \ \ \ \ \ \ \ \ \ \ \ \ \ \ \ \
\bigskip

\textbf{2\ \ \ Preons - metaflavors and metacolors \ \ \ \ \ \ \ \ \ \ \ \ \
\ \ \ \ \ \ \ \ \ \ \ \ \ \ \ \ \ \ \ \ \ \ \ \ \ \ \ \ \ \ \ \ \ \ \ \ \ \
\ \ \ \ \ \ \ \ \ } \ \ \ \ \ \ \ \ \ \ \ \ \ \ \ \ \ \ \ \ \ \ \ \ \ \ \ \
\ \ \ \ \ \ \ \ \ \ \ \ \ \ \ \ \ \ \ \ \ \ \ \ \ \ \ \ \ \ \ \ \ \ \ \ \ \
\ \ \ \ \ \ \ \ \ \ \ \ \ \ \ \ \ \ \ \ \ \ \ \bigskip

\textbf{3\ \ \ AM conditions for N metaflavors \ \ \ \ \ \ \ \ \ \ \ \ \ \ \
\ \ \ \ \ \ \ \ \ \ \ \ \ \ \ \ \ \ \ \ \ \ \ \ \ \ \ \ \ \ \ \ \ \ \ \ \ \
\ \ \ \ \ \ \ \ \ \ }

\ \ \ 3.1 \ Preamble \ \ \ \ \ \ \ \ \ \ \ \ \ \ \ \ \ \ \ \ \ \ \ \ \ \ \ \
\ \ \ \ \ \ \ \ \ \ \ \ \ \ \ \ \ \ \ \ \ \ \ \ \ \ \ \ \

\ \ \ 3.2 \ Strengthened AM condition: SU(8) unification

\ \ \ 3.3 \ Minimality matching condition \bigskip

\textbf{4 \ \ Composites - the L-R symmetry phase} \ \ \ \ \ \ \ \ \ \ \ \ \
\ \ \ \ \ \ \ \ \ \ \ \ \ \ \ \ \ \ \ \ \ \ \ \ \ \ \ \ \ \ \ \ \ \ \ \ \ \
\ \ \ \ \

\ \ \ 4.1 \ Multiplets of preon and composites\ \ \ \ \ \ \ \ \ \ \ \ \ \ \
\ \ \ \ \ \ \ \ \ \ \ \ \ \ \ \ \ \ \ \ \ \ \ \ \ \ \ \ \ \ \ \ \ \ \ \ \ \
\ \ \ \ \ \ \

\ \ \ 4.2 \ Structure of composite fermions \ \ \ \ \ \ \ \ \ \ \ \ \ \ \ \
\ \ \ \ \ \ \ \ \ \ \ \ \ \ \ \ \ \ \ \ \ \ \ \ \ \ \ \ \ \ \ \ \ \ \ \ \ \
\ \ \

\ \ \ 4.3 \ Composite scalars \ \ \ \ \ \ \ \ \ \ \ \ \ \ \ \ \ \ \ \ \ \ \
\ \ \ \ \ \ \ \ \ \ \ \ \ \ \ \ \ \ \ \ \ \ \ \ \ \ \

\ \ \ 4.4 \ Gravitational conversion of vectorlike fermions \bigskip

\textbf{5 \ \ Composites - partially broken L-R symmetry \ \ \ \ \ \ \ \ \ \
\ \ \ \ \ \ \ }\ \ \ \ \ \ \ \ \ \ \ \ \ \ \ \ \ \ \ \ \ \ \ \ \ \ \ \ \ \ \
\ \ \ \ \ \ \ \ \

\ \ \ 5.1 \ Strategy

\ \ \ 5.2 \ How may it work

\ \ \ 5.3 \ Metaflavor theory with broken $L$-$R$ symmetry

\ \ \ 5.4 \ Spectrum of survived composites \bigskip

\textbf{6 \ \ Some immediate physical consequences\ \ \ \ \ \ \ \ \ \ \ \ \
\ \ \ \ \ \ \ \ \ \ \ \ \ \ \ \ \ }

\ \ \ 6.1 \ Quarks, leptons and beyond\ \ \ \ \ \ \ \ \ \ \ \ \ \ \ \ \ \ \
\

\ \ \ 6.2 \ Quark-lepton masses and mixings

\ \ \ 6.3 \ Neutrino masses\

\ \ \ 6.4 \ Heavy states

\ \ \ \ 6.5 \ \ Basic scenarios\ \ \ \ \ \ \ \ \ \ \ \ \ \ \ \ \ \ \ \ \ \ \
\ \ \ \ \ \ \ \ \ \ \ \ \ \ \ \ \ \ \ \ \ \ \ \ \ \ \ \ \ \ \ \ \ \ \ \ \ \
\ \ \ \ \ \ \ \ \ \ \ \ \ \ \ \ \ \ \ \ \ \ \ \ \ \ \ \ \ \ \ \ \ \ \ \ \ \
\ \ \ \ \ \ \ \ \ \ \ \ \ \ \ \ \ \bigskip

\textbf{7 \ \ The family symmetry SU(3)}$_{F}$ \bigskip

\textbf{8 \ \ Conclusion and outlook}

\bigskip\bigskip\bigskip

\textbf{\ \ \ \ \ \ \ \ \ \ \ \ \ \ \ \ \ \ \ \ \ \ \ \ \ \ \ \ \ \ \ \ \ \
\ \ \ \ \ \ \ \ \ \ \ \ \ \ \ \ \ \ \ \ \ \ \ \ \ \ \ \ \ \ \ \ \ \ \ \ \ \
\ \ \ \ \ \ \ \ \ }

{\Large References}{\large \ \ }{\LARGE \ \ }\textbf{\ \ \ \ \ \ \ \ \ \ \ \
\ \ \ \ \ \ \ \ \ \ \ \ \ \ \ \ \ \ \ \ \ \ \ \ \ \ \ \ \ \ \ \ \ \ \ \ \ \
\ \ \ \ \ \ \ \ \ \ \ \ \ \ \ \ \ \ \ \ \ \ \ \ \ \ \ \ \ \ \ \ \ \ \ \ \ \
\ \ \ \ \ \ \ \ \ \ \ }

%
\thispagestyle{empty}\newpage

\section{Introduction}

As is well known, there is no a viable unification symmetry scheme for
classification of all observed quarks and leptons \cite{sl, moh}. Indeed,
while the known Grand Unified Theories, such as the $SU(5)$, $SO(10)$, or $%
E(6)$ GUTs, work well when applied in a supersymmetric context with all
three quark-lepton families included \cite{moh, st}, the very families are
taken mechanically or, at best, classified by their own (discrete, global or
local) symmetry. Any attempt to incorporate these families into a simple
grand unification symmetry group framework leads to high symmetries with
enormously extended representations which also contain lots of exotic states
that never been detected \cite{sl}. In fact, such states unavoidably appear
in any anomaly free combinations of representations of possible $SU(N)$ GUTs
with the $N$ value to be high enough to contain all three quark-lepton
families ($N\geq 8$). For the $SO(10)$ and $E(6)$ symmetry extensions the
situation looks even worse. Actually, for the $SO(4n+2)$ groups ($n\geq 3$),
which only have admissible complex representations, there are equal number $%
2^{2n-5}$ of the left-handed and right-handed spinor multiplets $16_{L,R}$
of the $SO(10)$ for quark-lepton families that requires an introduction of
some special survival mechanism in order to prevent them from pairing and
gaining the grand unification scale order masses. Meanwhile, a major
objection to the $E(6)$ symmetry extensions to the $E(7)$ and $E(8)$ which
could in principle include all three families of quarks and leptons is that
these extensions have only the self-conjugate representations that requires
some special (mostly unnatural) symmetry breaking pattern to provide the
chiral character of the weak interactions in the low-energy limit.

This problem may still motivate us to seek a solution in some subparticle or
preon model for quark and leptons, despite many interesting attempts made in
the past (some significant references can be found in \cite{moh, ds}). A
required solution might show, after all, which grand unification symmetry,
if any, could accommodate all the presumably composite quark and lepton
species. Some observed replications in their spectra, such as an existence
of several quark color triplets, as well as the quark and lepton weak
doublets, and as many as three identical quark-lepton families could mean
that there might exist the truly elementary fermions, preons, being actual
carriers of all the fundamental quantum numbers involved and composing the
observed quarks and leptons at larger distances. Generally, certain
regularities in replications of particles may signal about their constituent
structure. Indeed, just regularities in the spectroscopy of hadrons observed
in the nineteen-sixties made it possible to discover the quark structure of
hadrons in the framework of the so-called Eightfold Way. This term was
coined by Murray Gell-Mann in 1961 to describe a classification scheme for
hadrons, that he had devised, according to which the known baryons and
mesons are grouped into the eight-member families of some global hadron
flavor symmetry $SU(3)$ \cite{ew}. This had finally led to the hypothesis of
quarks which form the fundamental triplet of this symmetry, and,
consequently, to a compositeness of the observed baryons and mesons. We try
to show here that the inspirational Eightfold Way idea looks, as strange as
it may seem, much more relevant when it is applied to a next level of the
matter elementarity, namely, to elementary preons and composite quarks and
leptons. As we find below, just the eight preon model and their local
metaflavor symmetry $SU(8)_{MF}$ may under certain natural conditions
determine the fundamental entities of physical world and its total internal
symmetry at low energies.

Some heuristic motivation for such a picture may be related to the eight
fundamental quantum numbers (charges) presently observed. They correspond in
fact to the two weak isospin orientations, three types of colors and three
species of quark-lepton families. Their basic carriers could be eight
fermion preon fields $\mathcal{P}_{i}$ ($i=1,\ldots ,8$) being presumably
the fundamental octet of some basic\ metaflavor symmetry $SU(8)_{MF}$.
Accordingly, we will refer to these preons as a collection of "isons" $%
\mathcal{W}_{a}$ ($a=1,2$), "chromons" $\mathcal{C}_{k}$ ($k=1,2,3$) and
"famons" $\mathcal{F}_{p}$ ($p=1,2,3$). We find below that, apart from some
heuristic argumentation, the local metaflavor $SU(8)_{MF}$ symmetry as a
basic internal symmetry of the physical world at small distances is indeed
advocated by preon model for composite quarks and leptons.

We start by considering the $L$-$R$ symmetric composite model for quarks and
leptons where the $N$ left-handed and $N$\ right-handed constituent preons
possessing some vectorlike local metaflavor symmetry $SU(N)_{MF}$ are bound
by the chiral $SO(n)_{L}\times SO(n)_{R}$ gauge metacolor forces. They
generally possess an exact chiral global symmetry $SU(N)_{L}\times SU(N)_{R}$
as well, once their $SU(N)_{MF}$ gauge metaflavor interactions are switched
off. Though these interactions may in principle break the preon chiral
symmetry, they are typically too weak at the metacolor confining distances
to influence the bound state spectrum. This spectrum is proposed to only
depend on whether the preon chiral symmetry $SU(N)_{L}\times SU(N)_{R}$ or
some of its part is preserved at large distances. This in turn determines
the local metaflavor symmetry that could be observed at large distances
through the massless or low-lying fermion composites emerged. Any extra
metaflavor which is not protected by the preserved chiral symmetry lead to
the superheavy composites with the metacolor scale $\Lambda _{MC}$ order
masses and, therefore, appear unobserved at a laboratory scale.

We find that just the eight $L$-preons and eight $R$-preons and their
observed metaflavor symmetry $SU(8)_{MF}$ appear as a result of a solution
to the 't Hooft's anomaly matching condition \cite{t} providing the global
chiral $SU(8)_{L}\times SU(8)_{R}$ symmetry preservation and masslessness of
emerged composite fermions. We show that this happens if: (1) this condition
is satisfied individually for the $L$-preon and $R$-preon composites and (2)
each of these two series of composites fill the single irreducible
representations of the $SU(N)_{L}$ and $SU(N)_{R}$ symmetry groups,
respectively, rather than a set of their representations. Through the
constraints on the admissible chiral symmetry $SU(N)_{L}\times SU(N)_{R}$
providing the massless composite fermions at large distances, the anomaly
matching condition puts in general a powerful constraint on the underlying
local metaflavor symmetry $SU(N)_{MF}$ as well. However, such an emerged $L$-%
$R$ symmetric $SU(8)_{MF}$ metaflavor theory certainly appears pure
vectorlike for the identical $L$-preon and $R$-preon composite multiplets
involved. This means that, while preons are left massless being protected by
their own metacolors, the composites being metacolor singlets will pair up
and acquire heavy Dirac masses. We next show how the spontaneous $L$-$R$
symmetry violation caused by the simultaneously emerged composite scalars
reduces the initially vectorlike metaflavor $SU(8)_{MF}$ symmetry down to
one of its chiral subgroups being of significant physical interest.
Particularly, this violation implies that, while there still remains the
starting chiral symmetry $SU(8)_{L}$ for the left-handed preons and their
composites, for the right-handed states one may only have the broken chiral
symmetry $[SU(5)\times SU(3)]_{R}$. Therefore, whereas nothing really
happens with the left-handed preon composites still completing the total
multiplet of the $SU(8)_{L}$, the right-handed preon composites will form
only some particular submultiplets of the $[SU(5)\times SU(3)]_{R}$
symmetry. As a result, we eventually come from the $L$-$R$ symmetric
metaflavor $SU(8)_{MF}$ theory\ down to the conventional $SU(5)$ GUT
extended by an extra local family symmetry $SU(3)_{F}$ describing the three
standard families of quarks and leptons. Though the tiny confinement scale
for universal preons composing both quarks and leptons makes it impossible
to directly confirm their composite nature, simultaneous emergence of
several extra $SU(5)\times SU(3)_{F}$ multiplets of heavy composite fermions
may help with a model verification. Some of them through a natural see-saw
mechanism provide the physical neutrino masses which, in contrast to
conventional picture, appear to follow an inverted family hierarchy. Others
mix with ordinary quark-lepton families in a way that there may arise a
marked violation of unitarity in the CKM matrix for leptons depending on an
interplay between the compositeness scale and scale of the family symmetry $%
SU(3)_{F}$. All issues mentioned above are successively considered below in
the subsequent sections 2-7, and in the final section 8 we present our
conclusion. For simplicity, we work in an ordinary spacetime framework,
though extension to the conventional $N=1$ supersymmetry with preons and
composites treated as standard scalar superfields could generally be made.
As we see below, such an extension looks appealing in many aspects.

Some early attempt to classify quark-lepton families in the framework of the
$SU(8)$ GUT with composite quarks and leptons had been made quite a long ago
\cite{jp}, though with some special requirements which presently seem not
necessary or could be in principle derived rather than postulated. Since
then also many other things became better understood, especially the fact
that the chiral family symmetry $SU(3)_{F}$, taken by its own, was turned
out to be rather successful in description of quark-lepton families.
Meanwhile, as mentioned above, there has not yet appeared any valuable grand
unified symmetry scheme with simultaneous classification of all observed
quarks and leptons (though many interesting\ case studies have been carried
out \cite{z}). All that motivates us to address this essential problem once
again in the framework of a significantly modified preon model for composite
quarks and leptons. We will largely follow the recent paper \cite{npb}.

\section{Preons - metaflavors and metacolors}

We start formulating the key elements of the preon model considered here.
They are as follows below.

(1) We propose that at small distances there are $2N$ elementary massless
left-handed and right-handed preons, described by the independent Weil
spinors $P_{iL}$ and\ $Q_{iR}$ ($i=1,\ldots ,N$), which possess a common
local metaflavor symmetry $SU(N)_{MF}$ unifying all known physical charges,
such as weak isospin, color, and family number\footnote{%
Note that, unlike a conventional terminology, we call a metaflavor symmetry
a real local flavor symmetry of preons rather than a spectator metaflavor
symmetry appearing as result of a technical "gauging" of their global chiral
symmetry which we discuss below.}. The preons, both $P_{iL}$ and $Q_{iR}$,
transform under the fundamental representation of the $SU(N)_{MF}$ and their
metaflavor theory presumably has an exact $L$-$R$ symmetry. Actually, the $%
SU(N)_{MF}$\ appears at the outset as some vectorlike symmetry which then
breaks down at large distances to some of its chiral subgroup. All observed
quarks and leptons are proposed to consist of these universal preons some
combinations of which are collected in composite quarks, while others in
composite leptons.

(2) The preons also possess a local chiral metacolor symmetry $%
G_{MC}=G_{MC}^{L}\times G_{MC}^{R}$ which contains them in its basic
representation. In contrast to their common metaflavors, the left-handed and
right-handed preon multiplets being vectorlike under the $SU(N)_{MF}$
symmetry, are taken to be chiral under the metacolor symmetry $G_{MC}$. They
appear with different metacolors, $P_{iL}^{\boldsymbol{a}}$ and \ $Q_{iR}^{%
\boldsymbol{a}^{\prime }}$, where {$\boldsymbol{a}$} and $\boldsymbol{a}%
^{\prime }\ $are indices of the corresponding metacolor subgroups $%
G_{MC}^{L} $ ({$\boldsymbol{a}$ }${=1,...,n}$) and $G_{MC}^{R}${\ (}$%
\boldsymbol{a}^{\prime }$ ${=1,...,n}$), respectively. These chiral $%
G_{MC}^{L,R}$ metacolor forces bind preons into composites - quarks, leptons
and other states. As a consequence, there are two types of composites at
large distances being composed individually from the left-right and
left-handed preons, respectively. They presumably have a similar radius of
compositeness, $R_{MC}$ $\sim 1/\Lambda _{MC}$, where $\Lambda _{MC}$
corresponds to the scale of the preon confinement for the asymptotically
free (or infrared divergent) metacolor symmetries. Due to the proposed $L$-$%
R $ invariance, the metacolor symmetry groups $G_{MC}^{L}$ and $G_{MC}^{R}$
are taken identical with the similar scales for both of sets of preons. We
choose for metacolor the orthogonal symmetry group taking it, thereby, in a
chiral form
\begin{equation}
G_{MC}=SO(n)_{MC}^{L}\times SO(n)_{MC}^{R}\text{ \ }  \label{lrmc}
\end{equation}%
This \ choice has some advantages over the unitary metacolor symmetry
commonly used. Indeed, the orthogonal metacolor symmetry \cite{jp, bar} is
generically anomaly-free for its basic vector representation in which preons
are presumably located. Also, the orthogonal metacolor allows more possible
composite representations of the metaflavor symmetry $SU(N)_{MF}$ including
those which are described by\ tensors with mixed upper and lower metaflavor
indices.

(3) Under the local symmetries involved, both the metaflavor $SU(N)_{MF}$
and metacolor $G_{MC}$, the assignment of preons $P_{iL}^{\boldsymbol{a}}$
and \ $Q_{iR}^{\boldsymbol{a}^{\prime }}$ looks as
\begin{equation}
P_{iL}^{\boldsymbol{a}}(N,n,1)\text{ , \ }Q_{iR}^{\boldsymbol{a}^{\prime
}}(N,1,n)  \label{as}
\end{equation}%
In fact, they also possess an accompanying chiral global $SU(N)_{L}\times
SU(N)_{R}$ symmetry\footnote{%
Note that the $SU(N)_{L}\times SU(N)_{R}$ is a chiral symmetry of the
independent left-handed ($P_{iL}^{\boldsymbol{a}}$) and right-handed ($%
Q_{iR}^{\boldsymbol{a}^{\prime }}$) Weil spinors rather than chiral symmetry
related to $L$- and $R$-components of the same Dirac spinors, as is usually
implied. Instead of the $L$-$R$ basis used here, one could equally work in
the pure left-handed basis where one may properly have the preon assignment,
$P_{iL}^{\boldsymbol{a}}(N,n,1)$ and \ $\overline{Q}_{L}^{i\boldsymbol{a}%
^{\prime }}(\overline{N},1,n)$, possessing the chiral symmetry $%
SU(N)_{L}\times SU(N)_{L}^{\prime }$.} in the limit when their common gauge $%
SU(N)_{MF}$ metaflavor interactions are switched off. We omitted here the
Abelian chiral $U(1)_{L,R}$ symmetries since the corresponding currents have
Adler-Bell-Jackiw anomalies given by the triangle graphs with a pair of
metagluons \cite{th}. Accordingly, the current divergences for massless
preons $P_{iL}^{\boldsymbol{a}}$ and \ $Q_{iR}^{\boldsymbol{a}^{\prime }}$
are given by
\begin{equation}
\partial _{\mu }J_{L}^{\mu }=n\frac{g_{MC}^{2}}{16\pi ^{2}}F_{\mu \nu }^{[%
\boldsymbol{a,b}]}F_{\rho \sigma }^{[\boldsymbol{a,b}]}\epsilon ^{\mu \nu
\rho \sigma }\text{ , \ \ }\partial _{\mu }J_{R}^{\mu }=n\frac{%
g_{MC}^{\prime 2}}{16\pi ^{2}}F_{\mu \nu }^{\prime \lbrack \boldsymbol{a}%
^{\prime }\boldsymbol{,b}^{\prime }]}F_{\rho \sigma }^{\prime \lbrack
\boldsymbol{a}^{\prime }\boldsymbol{,b}^{\prime }]}\epsilon ^{\mu \nu \rho
\sigma }  \label{div}
\end{equation}%
where $F_{\mu \nu }^{[\boldsymbol{a,b}]}$ and $F_{\mu \nu }^{\prime \lbrack
\boldsymbol{a}^{\prime }\boldsymbol{,b}^{\prime }]}$ are the metagluon field
strengths being in adjoint representations of the $SO(n)_{MC}^{L,R}$ groups,
respectively, while $g_{MC}$ and $g_{MC}^{\prime }$ are the appropriate
gauge coupling constants\footnote{%
Note that all fields and other quantities related to the "right" metacolor
group $SO(n)_{MC}^{R}$ are denoted everywhere by the same letters as those
of the "left" metacolor group $SO(n)_{MC}^{L}$ but taken with a prime
symbol. The metacolor indices are given by two different sets of the bold
Latin letters: $\boldsymbol{a},\boldsymbol{b},\boldsymbol{c},\boldsymbol{d}%
,...$ for the $SO(n)_{MC}^{L}$ and $\boldsymbol{a}^{\prime },\boldsymbol{b}%
^{\prime },\boldsymbol{c}^{\prime },\boldsymbol{d}^{\prime }\mathbf{,...}$
for the $SO(n)_{MC}^{R}$. The metaflavor indices are given by ordinary
lowercase Latin letters, while lowercase Greek letters stand, as usual, for
conventional spacetime indices.}. Thus, the $U(1)_{L,R}$ symmetries which
might be associated with chiral preon number symmetries in the classical
Lagrangian, appear broken by the quantum corrections\footnote{%
Nevertheless, one could presumably still "feel" these symmetries at the
small distances, $r\ll $ $R_{MC}$, where the corrections (\ref{div}) may
become unessential due to asymptotic freedom in the metacolor theory
considered. One can refer to this regime as the valent preon approximation
in which one may individually recognize each preon in a composite fermion.}.
The point is, however, the chiral discrete subgroups $Z_{N}^{L,R}$ of the $%
U(1)_{L,R}$ symmetries are still preserved. Under $Z_{N}^{L,R}$ the preons
transform as%
\begin{equation}
P_{iL}^{\boldsymbol{a}}\rightarrow e^{i2\pi q_{L}^{(i)}/N}P_{iL}^{%
\boldsymbol{a}}\text{ , \ }Q_{iR}^{\boldsymbol{a}^{\prime }}\rightarrow
e^{i2\pi q_{R}^{(i)}/N}Q_{iR}^{\boldsymbol{a}^{\prime }}  \label{ds}
\end{equation}%
where the discrete charges $q_{L,R}^{(i)}$ are integers being defined modulo
$N$ only. They are taken to be $q_{L,R}^{(i)}=1$ for all $N$ preon species
being in the fundamental multiplet of the metaflavor $SU(N)_{MF}$. We call
them the chiral discrete preon numbers. The conditions for the $Z_{N}^{L,R}$
symmetries to be anomaly-free in the triangles $%
Z_{N}^{L,R}-SO(n)_{MC}^{L,R}-SO(n)_{MC}^{L,R}$ are given, respectively, by
the simple equations \cite{da}%
\begin{equation}
\sum_{i}^{N}q_{L,R}^{(i)}=0\text{ }mod\text{ }N
\end{equation}%
which are automatically satisfied. Note that chiral discrete symmetries $%
Z_{N}^{L}$ and $Z_{N}^{R}$ can be equally considered as the centers of the
groups $SU(N)_{L}$ and $SU(N)_{R}$, respectively. Therefore, the total
chiral symmetry of all the preons involved is in fact the symmetry
\begin{equation}
K(N)=SU(N)_{L}\times SU(N)_{R}  \label{ch}
\end{equation}%
which also includes its center identified as
\begin{equation}
C(N,N)=Z_{N}^{L}\times Z_{N}^{R}  \label{ch'}
\end{equation}%
We will refer to the latter, in what follows, as the chiral discrete preon
number symmetry.

(4) Obviously, the preon condensate $\left\langle \overline{P}%
_{L}Q_{R}\right\rangle $ which could cause the metacolor scale $\Lambda
_{MC} $ order masses for composites is principally impossible in the
left-right metacolor model (\ref{lrmc}). This is in sharp contrast to an
ordinary QCD case where the left-handed and right-handed quarks forms the $%
\left\langle \overline{q}_{L}q_{R}\right\rangle $ condensate thus leading to
the color scale $\Lambda _{C}$ order masses for composite mesons and
baryons. The fact that there is no the $\left\langle \overline{P}%
_{L}Q_{R}\right\rangle $ type condensate may be generally considered as a
necessary but not yet a sufficient condition for masslessness of composites.
The genuine massless fermion composites are presumably only those which
preserve chiral symmetry of preons (\ref{ch}) at large distances that is
controlled by the 't Hooft's anomaly matching (AM) condition \cite{t}. We
also consider here the composite scalar fields which could break the $%
SU(N)_{MF}$ metaflavor theory down to the Standard Model (SM). Generally,
they being no protected by any symmetry will become very heavy (with masses
of the order of the scale $\Lambda _{MC}$) and decouple from a low-lying
particle spectrum. However, candidate(s) for the composite SM Higgs boson
could also be expected under some special dynamical requirements.

(5) To summarize, we introduced $2N$ left-handed and right-handed preons, $%
P_{iL}^{\boldsymbol{a}}$ and\ $Q_{iR}^{\boldsymbol{a}^{\prime }}$ ($%
i=1,\ldots ,N$; {$\boldsymbol{a}=1,...,n$}; $\boldsymbol{a}^{\prime
}=1,...,n $), possessing some local metaflavor symmetry $SU(N)_{MF}$ and
chiral metacolor symmetry $SO(n)_{MC}^{L}\times SO(n)_{MC}^{R}$ with numbers
of metaflavors ($N$) and metacolors ($n$) not yet determined. We argued that
these preons also possess the chiral symmetry $K(N)$ (\ref{ch}) in the limit
when their gauge $SU(N)_{MF}$ metaflavor interactions are switched off.
Though these interactions break the chiral symmetry $K(N)$, we propose for
what follows that they are typically too weak at the metacolor confining
distances to influence the bound state spectrum. On the other hand, just a
preservation of the chiral symmetry $K(N)$ determines the particular
metaflavor symmetry $SU(N)_{MF}$ which could be observed at large distances
through the massless composites emerged. Indeed, any extra $\acute{N}$
metaflavors (extra $\acute{N}$ pairs of $P$ and $Q$ preons),\ which are not
provided by a preserved chiral symmetry, will form superheavy composites
with the metacolor scale $\Lambda _{MC}$ order masses and, therefore, appear
unobserved at a laboratory scale. If that would be the case, one might have
the chiral symmetry $K(N+\acute{N})$ for preons, while the lower symmetry $%
K(N)$ for composites. However, for simplicity's sake, we accept below that
just the preserved chiral symmetry $K(N)$ eventually determines an observed
metaflavor symmetry $SU(N)_{MF}$ so as to exclude any extra preons in the
theory. Note that, apart from preons being the actual carriers of both the
metaflavors and metacolors, there may also exist a number of elementary
chiral fermions only possessing the metacolors. We will refer to them as
"sterilons" which can be used for construction of composites as well. A
rather natural case could be sterilons being the gaugino (metagluino)
multiplets, $S_{L}^{[\boldsymbol{a,b}]}$ and $S_{R}^{\prime \lbrack
\boldsymbol{a}^{\prime }\boldsymbol{,b}^{\prime }]}$ of the $SO(n)_{MC}^{L}$
and $SO(n)_{MC}^{R}$, respectively, if one would properly supersymmetrize
the metacolor theory. Accordingly, their masslessness might be protected by
gauge invariance rather than the abovementioned chiral symmetry $K(N)$\ (\ref%
{ch}) being solely related to the massless preons $P_{iL}^{\boldsymbol{a}}$
and \ $Q_{iR}^{\boldsymbol{a}^{\prime }}$, respectively.

\section{AM conditions for N metaflavors}

\subsection{Preamble}

Usually, for tackling mathematical problems related to conservation of the
chiral symmetry $SU(N)_{L}\times SU(N)_{R}$, one turns it into the would-be
local symmetry group with some spectator gauge fields and fermions \cite{t}
being in fact mathematical tools only. This trick allows to properly analyze
the corresponding gauge anomaly cancellation thus checking the chiral
symmetry preservation for massless preons and composites at both small and
large distances. Though this symmetry is usually referred to as a metaflavor
symmetry, we will call it the spectator gauge symmetry. An actual metaflavor
theory in our model is an input local vectorlike $SU(N)_{MF}$\ symmetry
unifying $N$ left-handed and $N$\ right-handed preons, while their chiral
symmetry $SU(N)_{L}\times SU(N)_{R}$ is in fact global\footnote{%
This is somewhat similar to an interrelation between local and global
symmetries in the conventional quark model with the lightest $u$ and $d$
quarks where the actual local flavor theory is determined by the unified
electroweak symmetry $SU(2)_{W}\times U(1)_{Y}$ rather than their chiral
global symmetry $SU(2)_{L}\times SU(2)_{R}\times U(1)_{L+R}$.}. It is
important to see that, whereas in the $SU(N)_{MF}$ metaflavor theory gauge
anomalies of preons and composites are automatically cancelled out between
left-handed and right-handed states involved, in the spectator gauge $%
SU(N)_{L}\times SU(N)_{R}$ theory all anomalies have to be cancelled by
special multiplets of the metacolorless spectator fermions introduced
individually for the $SU(N)_{L}$ and $SU(N)_{R}$ sectors of the theory. As
was mentioned above, though the $SU(N)_{MF}$ metaflavor interactions may in
principle break the preon chiral symmetry (\ref{ch}), they are typically too
weak to influence the bound state spectrum.

In the proposed preon model the AM condition \cite{t}\ states in general
that the chiral $SU(N)_{L}^{3}\ $ and $SU(N)_{R}^{3}$ triangle anomalies
related to $N$ left-handed and $N$ right-handed preons have to match those
for massless composite fermions being produced by the $SO(n)_{MC}^{L}$ and $%
SO(n)_{MC}^{R}$ metacolor forces, respectively. Actually, fermions composed
from the left-handed preons and those composed from the right-handed ones
have to independently satisfy their own AM conditions. In contrast, in the
local $SU(N)_{MF}$ metaflavor theory being as yet vectorlike, the $%
SU(N)_{MF}^{3}$ metaflavor triangle anomalies of the $L$-preons and $R$%
-preons, as well as anomalies of their left-handed and right-handed
composites, will automatically compensate each other for any number $N$ of
the starting preon species. However, the AM condition through the
constraints on the admissible chiral symmetry $SU(N)_{L}\times SU(N)_{R}$
providing the masslessness of composite fermions at large distances, may put
in general a powerful constraint on this number, and thereby on the
underlying local metaflavor symmetry $SU(N)_{MF}$ itself as a potential
local symmetry of massless (or light) composites. This actually depends on
the extent to which the accompanying global chiral symmetry (\ref{ch}) of
preons remains at large distances.

In one way or another, the AM condition
\begin{equation}
\sum_{r}i_{r}a(r)=na(N)  \label{am}
\end{equation}%
for preons (the right side) and composite fermions (the left side) should be
satisfied. For the sake of brevity, the equation (\ref{am}) is
simultaneously written for both the left-handed and right handed preons and
their composites. Here $a(N)$ and $a(r)$ are the group coefficients of
triangle anomalies related to the groups $SU(N)_{L}$ or $SU(N)_{R}$ in (\ref%
{ch}) whose coefficients are calculated in an ordinary way, \ \ \ \
\begin{equation}
a(r)d^{ABC}=Tr(\{T^{A}T^{B}\}T^{C})_{r}\text{ }  \label{a}
\end{equation}%
where $T^{A}$ ($A,B,C=1,\ldots ,N^{2}-1$) are the $SU(N)_{L,R}$ generators
taken in the corresponding representation $r$. The $a(N)$ corresponds to a
fundamental representation and is trivially equal to $\pm 1$ (for
left-handed and right-handed preons, respectively), while $a(r)$ is related
to a representation $r$ for massless composite fermions. The value of the
factors $i_{r}$ give a number of times the representation $r$ appears in a
spectrum of composite fermions and is taken positive for the left-handed
states and negative for the right-handed ones. The anomaly coefficients for
composites $a(r)$ contain an explicit dependence on the number of preons $N$%
, due to which one could try to find this number from the AM condition taken
separately for the $L$- and $R$-preons and their composites. In general,
there are too many solutions to the condition (\ref{am}) for any value of $N$%
. Nevertheless, for some special, though natural, requirements an actual
solution may only appear for $N=8$, as we will see below.

Apart from the AM conditions (\ref{am}), there could be in general another
kind of constraint on composite models \cite{t}. This constraint requires
the anomaly matching condition to work, even if any number of the initially
introduced $N$ preons acquire an infinite masses and consequently decouple
from an entire theory. Indeed, this extra constraint makes the AM conditions
to be independent of a number of metaflavors $N$ and, as a result,
classification of composite fermions becomes quite arbitrary. The point is,
however, that in our preon model with a chiral metacolor symmetry $%
SO(n)_{MC}^{L}\times SO(n)_{MC}^{R}$ the left-handed and\ right-handed
preons might only form the Majorana masses which are in fact forbidden by
the input local metaflavor symmetry $SU(N)_{MF}$ of the theory. Apart from
that, such a constraint may appear generally irrelevant since, as can be
seriously argued \cite{moh}, the nonperturbative effects may not be analytic
in the preon mass so that the theories for massless and massive preons are
turned out to be quite different. One way or the other, we will not consider
this extra constraint in our composite model.

\subsection{Strengthened AM condition: SU(8) unification}

To strengthen the AM condition one may think that it would be more
appropriate to have all composite quarks and leptons in a single
representation of the unified symmetry group $SU(N)_{MF}$ rather than in
some set of its representations. This, though would not largely influence
the gauge sector of the unified theory, could make its Yukawa sector much
less arbitrary. Apart from that, the composites belonging to different
representations would have in general the different discrete $Z_{N}^{L,R}$\
preon numbers that could look rather unnatural. In this manner, the
strengthened AM condition suggests that only some particular $SU(N)_{MF}$\
multiplets of the left-handed and right-handed composites being in the
corresponding representations of chiral symmetry groups $SU(N)_{L,R}\ $in (%
\ref{ch}) may contribute. It is clear that these $L$- and $R$-multiplets
have to be similar to remain the $SU(N)_{MF}$ metaflavor theory anomaly
free. Let us propose for the moment that we only have the minimal
three-preon fermion composites. They are formed by the metacolor forces
which correspond to the $SO(3)_{MC}^{L}\times SO(3)_{MC}^{R}$ symmetry case
in (\ref{lrmc}) with metagluons, $A_{\mu }^{\boldsymbol{a}}$ and $%
\boldsymbol{A}_{\mu }^{\prime \boldsymbol{a}^{\prime }}$ transforming, like
preons $P_{iL}^{\boldsymbol{a}}$ and $Q_{iR}^{\boldsymbol{a}^{\prime }}$
themselves, according to the vector representations of the $SO(3)_{MC}^{L}$
and $SO(3)_{MC}^{R}$, respectively. We will, therefore, require that only
some single representation $r_{0}$ for massless three-preon states has to
satisfy the AM condition that simply gives in (\ref{am})%
\begin{equation}
a(r_{0})=3\text{ }  \label{am1}
\end{equation}%
individually for $L$-preon and $R$-preon composites.

Now, calculating the anomaly coefficients for all possible three $L$-preon
and three $R$-preon metacolorless composites one has, respectively,
\begin{eqnarray}
&&\Psi _{\{ijk\}L}\text{ , }\Psi _{\{ijk\}R}^{\prime }\text{ , }\pm
(N^{2}/2+9N/2+9),  \notag \\
&&\Psi _{\lbrack ijk]L,R}\text{ , }\Psi _{\lbrack ijk]R}^{\prime }\text{ , }%
\pm (N^{2}/2-9N/2+9),  \notag \\
&&\Psi _{\{[ij]k\}L}\text{ , }\Psi _{\{[ij]k\}R}^{\prime }\text{ , }\pm
(N^{2}-9),  \notag \\
&&\Psi _{\{jk\}L}^{i}\text{ , \ }\Psi _{\{jk\}R}^{\prime i}\text{ , }\pm
(N^{2}/2+7N/2-1),  \notag \\
&&\Psi _{\lbrack jk]L}^{i}\text{ , }\Psi _{\lbrack jk]R}^{\prime i}\text{ ,
\ }\pm (N^{2}/2-7N/2-1)  \label{tens}
\end{eqnarray}%
where all appropriate third-rank \ representations of the $SU(N)_{L,R}$ are
listed (the traces have been subtracted from the tensors at the last two
lines)\footnote{%
The anomaly coefficients\ for all possible third-rank tensors given above
have been first calculated in \cite{jp, bar}).}. According to the notation
taken$^{3}$, we denote the $R$-preon composites by the same letters as the $%
L $-preon ones but taken with the prime symbol (anomaly coefficients for
right-handed composites have to be taken with an opposite sign). In contrast
to conventional QCD, for an orthogonal metacolor there could also exist the
simplest composite states $\Psi _{iL}\ $and $\Psi _{iR}^{\prime }$
constructed out of single $P_{iL}^{\boldsymbol{a}}$ and $Q_{iR}^{\boldsymbol{%
a}^{\prime }}$ preons, whose metacolor charges are screened by the
corresponding metagluon fields of $SO(3)_{MC}^{L}$ and $SO(3)_{MC}^{R}$,
respectively. Though they could in principle contribute into the AM
condition, we do not consider these fundamental $SU(8)_{MF}$ multiplets as
appropriate composite candidates for physical quarks and leptons. In our
left-right model the composite states $\Psi _{iL}\ $and $\Psi _{iR}^{\prime
} $ generally pair up and receive heavy Dirac masses (as is argued later).

Putting now each of the above anomaly coefficients in (\ref{tens}) into the
AM condition (\ref{am1}) one can readily find that there is a solution with
an integer $N$ only for the last tensors $\Psi _{\lbrack jk]L}^{i}$ and $%
\Psi _{\lbrack jk]R}^{\prime i}$, and this is in fact the unique "eightfold"
solution
\begin{equation}
N^{2}/2-7N/2-1=3,\text{ }N=8\text{ .}  \label{8-1}
\end{equation}%
This means that among all possible chiral symmetries $K(N)$ in (\ref{ch})
only $SU(8)_{L}\times SU(8)_{R}$ could in principle provide the massless
fermion composites at large distances that in turn identifies the metaflavor
$SU(8)_{MF}$ symmetry as a possible unified symmetry of the left-handed and
right-handed preons and their composites which presumably include, among
many other states, the ordinary quarks and leptons.

Remarkably, the same solution $N=8$ independently appears if one requires
that the yet unknown metaflavor symmetry $SU(N)_{MF}$ has to possess the
right $SU(5)$ GUT classification \cite{gg} for the observed quarks and
leptons. Indeed, for an adequate $SU(5)$ assignment the "right" $SU(N)_{MF}$
representation has to contain the equal numbers of the $SU(5)$ anti-quintets
$\overline{5}$ and decuplets $10$ to be in accordance with the Standard
Model classification and an observation. Decomposing $SU(N)_{MF}$ into $%
SU(5)\times $ $SU(N-5)$ one can calculate these numbers for all its $3$%
-index multiplets presented in (\ref{tens}). Actually, as can be easily
seen, neither of them but the representations $\Psi _{\lbrack jk]L}^{i}$ and
$\Psi _{\lbrack jk]R}^{\prime i}$ appear to contain both chiral
anti-quintets and decuplets of the $SU(5)$ GUT. Moreover, an equality of
their numbers in these representations determines the total symmetry group
itself for which such an equality is solely possible. In fact, this equality
when being required for the multiplets $\Psi _{\lbrack jk]L}^{i}$ and $\Psi
_{\lbrack jk]R}^{\prime i}$ reads as%
\begin{equation}
(N-5)(N-6)/2=N-5\text{ }  \label{8}
\end{equation}%
with the number of anti-quintets on the left side and number of decuplets on
the right one. As a result, we come once again to the eightfold $SU(8)_{MF}$
metaflavor symmetry ($N=8$).

Let us note, however, that the $SO(3)_{MC}^{L,R}$ metacolors providing the
three-preon structure of composite quarks and leptons may appear
insufficient for the preon confinement, unless one invokes some special
strong coupling regime \cite{wil1}. Generally, for $N$ preons possessing the
asymptotically free $SO(n)$ metacolor symmetry, one must require the
inequality \cite{wil}%
\begin{equation}
n>2+2N/11  \label{cf}
\end{equation}%
as a necessary condition for their confinement inside of quarks and leptons.
One can readily see that for a commonly used minimal metacolor number value,
$n=3$, at most five preon species ($N=5$) with an underlying metaflavor
symmetry $SU(5)_{MF}$ is only admissible. This certainly seems to be not
enough for all elementary quantum numbers (or metaflavors) presently
observed including those for quark-lepton families. For the next odd number,
$n=5$, one may have up to sixteen ($N=16$) admissible preon species that
could be quite sufficient. We consider here just this case in the
strengthened AM\ condition (\ref{am1}). Thus, the metacolor forces now
correspond to the $SO(5)_{MC}^{L}\times SO(5)_{MC}^{R}$ symmetry case with
gauge metagluon fields $A_{\mu }^{[\boldsymbol{a},\boldsymbol{b}]}$ and $%
A_{\mu }^{\prime \lbrack \boldsymbol{a}^{\prime },\boldsymbol{b}^{\prime }]}$
transforming according to adjoint representations of the $SO(5)_{MC}^{L}$
and $SO(5)_{MC}^{R}$ groups, respectively. Remarkably, due to an orthogonal
nature of metacolor, that allows a metacolor charge to be screened by its
own metagluons, hand in hand with the five-preon fermion composites, the
three-preon composites (\ref{tens}), as well as the one-preon composites,
may also appear in the bound spectrum in this case. We consider here just
these minimal composite states in the new strengthened AM\ condition%
\begin{equation}
a(r_{0})=5\text{ }  \label{am2}
\end{equation}%
individually for $L$-preon and $R$-preon composite states ignoring the
five-preon states as somewhat exotic ones.

Checking all the representations (\ref{tens}) of the $SU(N)_{L,R}$ we find
that the AM condition works again for the multiplets $\Psi _{\lbrack
jk]L}^{i}$ and $\Psi _{\lbrack jk]R}^{\prime i}$ provided that they are
taken together with the fundamental multiplets $\Psi _{iL}^{(1,2)}$ and $%
\Psi _{iR}^{\prime (1,2)}$ corresponding to composites in which the preon
metacolor is properly screened by a pair of metagluons\footnote{%
Or, alternatively, one can take $\Psi _{iL}^{(1)}$ and $\Psi _{iR}^{\prime
(1)}$\ as the single preon states screened by a pair of metagluons, while $%
\Psi _{iL}^{(2)}$ and $\Psi _{iR}^{\prime (2)}$ as the actual three-preon
states given by the "trace" tensors obtained after taking traces out of the
proper three-index tensors $\Psi _{\lbrack jk]L}^{i}$ and $\Psi _{\lbrack
jk]R}^{\prime i}$\ in (\ref{tenss}). Their wave functions can be found in
the section 4.2.},%
\begin{equation}
\Psi _{\lbrack jk]L}^{i}+\Psi _{iL}^{(1)}+\Psi _{iL}^{(2)}\text{ , \ }\Psi
_{\lbrack jk]R}^{\prime i}+\Psi _{iR}^{\prime (1)}+\Psi _{iR}^{\prime (2)}%
\text{\ .}  \label{tenss}
\end{equation}%
Again, using the AM condition (\ref{am2}) for the tensors (\ref{tenss}) one
eventually has the same equation\ (\ref{8-1}) for the number of preon
species $N$, thus coming once again to the eightfold solution $N=8$.
Generally, due to direct screening effects of the orthogonal metacolor the
massless three-preon and one-preon composites mentioned above may appear for
any odd number of metacolors rather than in the $n=5$ metacolor case only.
Indeed, one can easily check that for any higher odd $n$ value the AM
condition also works for the three-preon tensors $\Psi _{\lbrack jk]L}^{i}$ (%
$\Psi _{\lbrack jk]R}^{\prime i}$) extended by some number $p$ of one-preon
multiplets $\Psi _{iL}$ ($\Psi _{iR}^{\prime }$)
\begin{equation}
\Psi _{\lbrack jk]L}^{i}+p\Psi _{iL}\text{ , \ }\Psi _{\lbrack jk]R}^{\prime
i}+p\Psi _{iR}^{\prime }\text{ }  \label{tensss}
\end{equation}%
provided that they all are properly screened by the corresponding
metagluons. Actually, the AM condition now leads to the equation
generalizing the above anomaly matching condition (\ref{8-1})%
\begin{equation}
N^{2}/2-7N/2-1+p=n  \label{gam}
\end{equation}%
One can see that there appear solutions in (\ref{gam}) only for $n-p=3$ and,
therefore, one has again solution for the eightfold chiral symmetry $%
SU(8)_{L}\times SU(8)_{R}$ and, thereby, the eightfold metaflavor symmetry $%
SU(8)_{MF}$. Indeed, the $n=5$ metacolor case leading to the composite
multiplets (\ref{tenss}) is just one particular, though minimal confining
model case which satisfies the generalized anomaly matching condition (\ref%
{gam}).

\subsection{Minimality matching condition}

We have seen above that for a minimal metacolor case, $n=3$, the
strengthened AM condition (\ref{am1}) determines the $SU(8)_{MF}$ metaflavor
symmetry as the only possible unified symmetry of massless preons and their
composites which are located in the representations $\Psi _{\lbrack
jk]L}^{i} $ and $\Psi _{\lbrack jk]R}^{\prime i}$. Nonetheless, for the
higher numbers of metacolors some other candidates may also emerge. Thanks
to the orthogonal metacolor screening effect, we may consider again only
three-preon composite tensors ignoring the high-number preon states as some
exotic ones. One can immediately check that there appears a new candidate
even in the five-metacolor case. Indeed, the strengthened AM condition (\ref%
{am2}) for the third-rank antisymmetric composite multiplets in (\ref{tens})
\begin{equation}
\Psi _{\lbrack ijk]L}\text{ , }\Psi _{\lbrack ijk]R}^{\prime }\text{ ; \ }%
N^{2}/2-9N/2+9=5  \label{82}
\end{equation}%
has a solution for an integer $N$ and this is again the eightfold solution, $%
N=8$. Therefore, we have in fact two competing eightfold models conditioned
by the survived $SU(8)_{L}\times SU(8)_{R}$ chiral symmetry at large
distances. Meanwhile, only the model with composite multiplets $\Psi
_{\lbrack jk]L}^{i}$ and $\Psi _{\lbrack jk]R}^{\prime i}$ may give, as was
clearly seen above, an adequate description of the quark-lepton families.

Some way to discriminate the model with composite multiplets (\ref{82}), as
well as models with other three-preon tensors in (\ref{tens}) would be an
existence of an extra selection rule related to minimality of possible
global preon numbers for composites, both left-handed and right-handed ones.
Since, as we discussed in the section 2 above, the corresponding symmetries $%
U(1)_{L,R}$ are broken by anomalies$^{4}$, one may only turn to the discrete
preon numbers related to the center $Z_{N}^{L}\times Z_{N}^{R}$ of the
chiral $SU(N)_{L}\times SU(N)_{R}$ symmetry involved. So, according to this
selection rule only composites satisfying the discrete preon number matching
condition%
\begin{equation}
Z_{N}^{L,R}(preons)=Z_{N}^{L,R}(composites)  \label{yy}
\end{equation}%
might presumably appear in physical spectrum. We cannot fundamentally argue
why the extra selection rule (\ref{yy}) should work in general. However, it
seems reasonable that the orthogonal metacolor allowing (in contrast to
unitary metacolor symmetry case) too many possible composite configurations
may somehow dynamically single out composites with a minimal discrete preon
number.

If the minimality matching condition (\ref{yy}) works, then the composite
multiplets in (\ref{tenss}) and \ref{tensss}) having the minimal unit $%
Z_{N}^{L,R}$ charges (as those of the preons themselves) may emerge massless
at large distances, while the composite third-rank multiplets (\ref{82})
having the triple discrete charges should presumably acquire heavy masses.
In general, only one more multiplet of massless composites with minimal $%
Z_{N}^{L,R}$ charges, namely $\Psi _{\{jk\}L}^{i}$ ($\Psi _{\{jk]\}}^{\prime
i}$), could in principle appear in the collection (\ref{tens}). However, as
one can readily check, the AM condition for this multiplet would require an
enormously large number of metacolors $n$ even for a low number of
metaflavor species $N$ that seems hardly admissible.

So, we will mainly consider in what follows the set of multiplets (\ref%
{tenss}) satisfying the AM conditions in the $n=5$ metacolor case that
eventually leads to the $SU(8)_{MF}$ metaflavor symmetry which may contain
three standard families of quarks and leptons. Generally, the whole class of
the $n$ metacolor $SU(8)_{MF}$ theories emerging through the massless
composite multiplets (\ref{tensss}) appears the best possible choice, though
one has still vectorlike theory with similar left-handed and right-handed
composite multiplets. Remarkably, this class is also selected by the
discrete preon number matching condition (\ref{yy}) discussed above. We find
in the section 5 below that the proposed selection rule (\ref{yy}) becomes
particularly important when the starting $L$-$R$ symmetry in the theory is
spontaneously broken and, as a result, three chiral families of quarks and
leptons emerge.

\section{Composites - the L-R symmetry phase}

\subsection{Multiplets of preon and composites}

According to the strengthened AM conditions discussed above only the chiral
symmetry (\ref{ch}) of eight preon species
\begin{equation}
K(8)=SU(8)_{L}\times SU(8)_{R}\text{ \ }  \label{881}
\end{equation}%
with the center%
\begin{equation}
C(8,8)=Z_{8}^{L}\times Z_{8}^{R}  \label{881'}
\end{equation}%
may appear unbroken at both small and large distances. This in turn selects,
among many other $SU(N)_{MF}$ alternatives, just the local metaflavor $%
SU(8)_{MF}$ theory as the only possible unified theory of massless preons
and composites. So, we have at small distances the sixteen left-handed and
right-handed preons given by the Weil fields
\begin{equation}
P_{iL}^{\boldsymbol{a}}\text{ , \ }Q_{iR}^{\boldsymbol{a}^{\prime }}\text{ \
\ \ \ }(i=1,\ldots ,8;\text{ }{\boldsymbol{a}}=1,...,5;\text{ }\boldsymbol{a}%
^{\prime }=1,...,5)  \label{88}
\end{equation}%
belonging to the fundamental octets and quintets of the $SU(8)_{MF}$
symmetry and metacolor symmetry $SO(5)_{MC}^{L}\times SO(5)_{MC}^{R}$,
respectively. At large distances, on the other hand, we have massless
composites (\ref{tenss}) which are located in the left-handed and
right-handed multiplets of the $SU(8)_{MF}$
\begin{equation}
216_{[jk]L}^{i}+8_{iL}^{(1)}+8_{iL}^{(2)}\text{ , \ }%
216_{[jk]R}^{i}+8_{iR}^{(1)}+8_{iR}^{(2)}\text{ ,}  \label{216}
\end{equation}%
where their dimensions are explicitly indicated. Recall that, while the
three-preon composites $216_{L,R}$ are properly screened by single
metagluons of the $SO(5)_{MC}^{L,R}$, the one-preon composites $%
8_{L,R}^{(1,2)}$ are screened by pairs of metagluons.

Note that the\ strengthened AM conditions propose that the chiral symmetry
subgroups $SU(8)_{L}\ $and $SU(8)_{R}$ of the left-handed and right-handed
preons in (\ref{881}) have to remain individually. At the same time, in the
vectorlike metaflavor $SU(8)_{MF}$\ theory these $L$- and $R$-preons act
jointly, so that all triangle anomalies at both small and large distances
always appear automatically compensated. Decomposing the $SU(8)_{MF}$\
composite multiplets (\ref{216}) into the $SU(5)\times SU(3)$ components one
has%
\begin{eqnarray}
216_{L,R} &=&\left[ (\overline{5}+10,\text{ }\overline{3}%
)+(45,1)+(5,8+1)+(24,3)+(1,3)+(1,\overline{6})\right] _{L,R}  \notag \\
8_{L,R}^{(1,2)} &=&\left[ (5,1)+(1,3)\right] _{L,R}^{(1,2)}  \label{216'}
\end{eqnarray}%
where the first term for the left-handed composites in $216_{L}$, $(%
\overline{5}+10,$ $\overline{3})_{L}$, could be associated with the standard
$SU(5)$ GUT\ assignment for quarks and leptons \cite{gg} extended by some
family symmetry, which we will denote for what follows by $SU(3)_{F}$. The
other submultiplets in (\ref{216'}) become heavy, as we show later, and
decouple from an observed low-lying particle spectrum. Below, we discuss in
more detail some possible $L$-preon and $R$-preon composites, both fermions
and scalars, being of an important interest.

\subsection{Structure of composite fermions}

The determination of an explicit form of wave functions for the composite
states (\ref{216}) is a complicated dynamical problem related to the yet
unknown dynamics of the preon confinement. We propose that some basic
features of the left-handed composites built from the $P$-preons\ and their
metagluons may be simply given by the gauge invariant expressions
\begin{eqnarray}
\text{ }\Psi _{\lbrack jk]L}^{i}(x) &\propto &\epsilon _{\boldsymbol{abcde}}%
\left[ \left( \overline{P}_{L}^{\boldsymbol{a}i}\gamma ^{\mu }D^{\nu
}P_{jL}^{\boldsymbol{b}}\right) P_{kL}^{\boldsymbol{c}}\right] F_{\mu \nu
}^{[\boldsymbol{de}]}\text{, \ }  \notag \\
\Psi _{iL}(x) &\propto &\epsilon _{\boldsymbol{abcde}}P_{iL}^{\boldsymbol{a}%
}F_{\mu \nu }^{[\boldsymbol{bc}]}F_{\mu \nu }^{[\boldsymbol{de}]}\text{\ }
\label{wf}
\end{eqnarray}%
for the three-preon and one-preon states, respectively ($F_{\mu \nu }^{[%
\boldsymbol{de}]}$ is the metagluon stress tensor related to the metacolor
symmetry $SO(5)_{MC}^{L}$ with the antisymmetric fifth-rank tensors $%
\epsilon _{\boldsymbol{abcde}}$ and vector representation indices $%
\boldsymbol{a}$, $\boldsymbol{b}$, $\boldsymbol{c}$, $\boldsymbol{d}$ and $%
\boldsymbol{e}$). The preon covariant derivative $D^{\nu }P_{jL}^{%
\boldsymbol{b}}$ in (\ref{wf}) should contain terms corresponding to both
metacolor and metaflavor local symmetries involved%
\begin{equation}
D^{\nu }P_{jL}^{\boldsymbol{b}}=\left[ \delta _{\boldsymbol{c}}^{\boldsymbol{%
b}}\delta _{j}^{l}\partial ^{\nu }+g_{MC}(I\cdot \mathbf{A}^{\nu })_{%
\boldsymbol{c}}^{\boldsymbol{b}}\delta _{j}^{l}+g_{MF}\delta _{\boldsymbol{c}%
}^{\boldsymbol{b}}(T\cdot \mathbf{B}^{\nu })_{j}^{l}\right] P_{lL}^{%
\boldsymbol{c}}  \label{cov}
\end{equation}%
where $I$ and $T$ stand for sets of the $SO(5)_{MC}^{L}$ and $SU(8)_{MF}$
generators, while $\mathbf{A}^{\nu }$ and $\mathbf{B}^{\nu }$ are the
corresponding gauge field multiplets taken with their coupling constants. In
the valent preon approximation, the left-handed preon currents (\ref{wf})
correspond to the zero mass bound states with a spin of $1/2$ and a helicity
$-1/2$ being formed by two left-handed preons and one anti-preon plus a
metagluon (or one left-handed preon plus a pair of metagluons) which are
moving in a common direction.

In a similar way one can construct the right-handed preon composites which
correspond to multiplets of states with a spin of $1/2$ and helicity $+1/2$
composed from right-handed $Q$-preons and metagluons of the metacolor
symmetry $SO(5)_{MC}^{R}$. This is simply achieved by making the proper
replacements in (\ref{wf}) leading to the composite states
\begin{eqnarray}
\Psi _{\lbrack jk]R}^{\prime i}(x) &\propto &\epsilon _{\boldsymbol{a}%
^{\prime }\boldsymbol{b}^{\prime }\boldsymbol{c}^{\prime }\boldsymbol{d}%
^{\prime }\boldsymbol{e}^{\prime }}\left[ \left( \overline{Q}_{R}^{%
\boldsymbol{a}^{\prime }i}\gamma ^{\mu }D^{\nu }Q_{jR}^{\boldsymbol{b}%
^{\prime }}\right) Q_{kR}^{\boldsymbol{c}^{\prime }}\right] F_{\mu \nu
}^{\prime \lbrack \boldsymbol{d}^{\prime }\boldsymbol{e}^{\prime }]}  \notag
\\
\Psi _{iR}^{\prime }(x) &\propto &\epsilon _{\boldsymbol{a}^{\prime }%
\boldsymbol{b}^{\prime }\boldsymbol{c}^{\prime }\boldsymbol{d}^{\prime }%
\boldsymbol{e}^{\prime }}Q_{iR}^{\boldsymbol{a}^{\prime }}F_{\mu \nu
}^{\prime \lbrack \boldsymbol{b}^{\prime }\boldsymbol{c}^{\prime }]}F_{\mu
\nu }^{\prime \lbrack \boldsymbol{d}^{\prime }\boldsymbol{e}^{\prime }]}
\label{wf1}
\end{eqnarray}%
where the antisymmetric fifth-rank tensors $\epsilon _{\boldsymbol{a}%
^{\prime }\boldsymbol{b}^{\prime }\boldsymbol{c}^{\prime }\boldsymbol{d}%
^{\prime }\boldsymbol{e}^{\prime }}$ and vector representation indices $%
\boldsymbol{a}^{\prime }$, $\boldsymbol{b}^{\prime }$, $\boldsymbol{c}%
^{\prime }$and $\boldsymbol{d}^{\prime }$ now belong to the metacolor
symmetry $SO(5)_{MC}^{R}$ with a metagluon $A_{\mu }^{\prime \lbrack
\boldsymbol{d}^{\prime }\boldsymbol{e}^{\prime }]}$. Accordingly, the
covariant derivative $D^{\nu }Q_{jR}^{\boldsymbol{b}^{\prime }}$ has a form%
\begin{equation}
D^{\nu }Q_{jR}^{\boldsymbol{b}^{\prime }}=\left[ \delta _{\boldsymbol{c}%
^{\prime }}^{\boldsymbol{b}^{\prime }}\delta _{j}^{l}\partial ^{\nu
}+g_{MC}^{\prime }(I^{\prime }\cdot \mathbf{A}^{\prime \nu })_{\boldsymbol{c}%
^{\prime }}^{\boldsymbol{b}^{\prime }}\delta _{j}^{l}+g_{MF}\delta _{%
\boldsymbol{c}^{\prime }}^{\boldsymbol{b}^{\prime }}(T\cdot \mathbf{B}^{\nu
})_{j}^{l}\right] Q_{lR}^{\boldsymbol{c}^{\prime }}  \label{cov'}
\end{equation}%
where the $SO(5)_{MC}^{R}$ set of generators $I^{\prime }$ and the
corresponding gauge field multiplet $\mathbf{A}^{\prime \nu }$ with its
coupling constant $g_{MC}^{\prime }$ are now presented.

In principle, apart from the composites (\ref{wf}) and (\ref{wf1}), one
might construct the composites with inverse chiralities, particularly, the
right-handed multiplets $\Psi _{\lbrack jk]R}^{i}$ and $\Psi _{iR}$\
composed from the left-handed preons $P_{iL}^{\boldsymbol{a}}$ and\ the
left-handed multiplets $\Psi _{\lbrack jk]L}^{\prime i}$ and $\Psi
_{iL}^{\prime }$\ composed from the right-handed preons $Q_{iR}^{\boldsymbol{%
a}^{\prime }}$. If so, they might pair up with the starting multiplets (\ref%
{wf}) and (\ref{wf1}) and, thereby, cause the metacolor scale $\Lambda _{MC}$
order masses individually for the $P$- and $Q$-preon composites. The point
is, however, that due the chiral symmetry preservation, according to which
the AM condition selects just the multiplets (\ref{wf}) and \ref{wf1}) as
the generically massless ones, the inverse chirality multiplets mentioned
above are not allowed to appear in the composite spectrum. In contrast, all
other composite fermion multiplets listed above in (\ref{tens}) may possess
both chiralities individually for the $P$- and $Q$-preon composites. As\ a
result, they will pair up and acquire the metacolor scale order Dirac
masses. In fact, all of them then decouples from the low-lying fermion
spectrum and may only contribute in the coupling running at superhigh
energies.

\subsection{Composite scalars}

Let us now turn to the heavy scalar composites which may cause spontaneous
breaking of the $L$-$R$ symmetry in the theory, as well as the metaflavor
symmetry $SU(8)_{MF}$ itself down to the Standard model and finally its
breaking as well. We schematically present below some of them. First, there
are the two-preon composites giving the massive effective scalar states
which complete in general both two-index symmetric and antisymmetric
multiplets of the $SU(8)_{MF}$%
\begin{eqnarray}
\chi _{\lbrack ij]}(x) &\propto &P_{[iL}^{\boldsymbol{a}}CP_{jL]}^{%
\boldsymbol{a}}\text{ , \ }\chi _{\{ij\}}(x)\propto P_{\{iL}^{\boldsymbol{a}%
}CP_{jL\}}^{\boldsymbol{a}}  \notag \\
\chi _{\lbrack ij]}^{\prime }(x) &\propto &Q_{[iR}^{\boldsymbol{a}^{\prime
}}CQ_{jR]}^{\boldsymbol{a}^{\prime }}\text{ , \ }\chi _{\{ij\}}^{\prime
}(x)\propto Q_{\{iR}^{\boldsymbol{a}^{\prime }}CQ_{jR\}}^{\boldsymbol{a}%
^{\prime }}  \label{s2}
\end{eqnarray}%
where $C$ stands for the charge-conjugation matrix. They have the following $%
SU(5)\times SU(3)_{MF}$ decomposition%
\begin{eqnarray}
28 &=&(5,3)+(10,1)+(1,\overline{3})\text{ ,}  \notag \\
36 &=&(5,3)+(15,1)+(1,6)  \label{s22}
\end{eqnarray}%
according to which they might break the family symmetry $SU(3)_{MF}$ when
developing vacuum expectation values (VEVs) on the components $(1,\overline{3%
})$ and $(1,6)$, respectively.

Next is the preon-antipreon composites belonging to the adjoint
representations of the $SU(8)_{MF}$%
\begin{equation}
\phi _{j}^{i}(x)\propto \overline{P}_{L}^{\boldsymbol{a}i}\gamma _{\mu
}D^{\mu }P_{jL}^{\boldsymbol{a}}\text{ , \ }\phi _{j}^{\prime i}(x)\propto
\overline{Q}_{R}^{\boldsymbol{a}i}\gamma _{\mu }D^{\prime \mu }Q_{jR}^{%
\boldsymbol{a}}  \label{s3}
\end{equation}%
where $D^{\mu }$ and $D^{\prime \mu }$ are the covariant derivatives with
respect to the metacolor and metaflavor symmetries given above in (\ref{cov}%
) and (\ref{cov'}), respectively. The $SU(5)\times SU(3)_{MF}$ structure of
these multiplets
\begin{equation}
63=(24,1)+(\overline{5},3)+(5,\overline{3})+(1,8)+(1,1)  \label{s33}
\end{equation}%
shows that they can be used for the "diagonal" breaking all the symmetries
involved - $SU(8)_{MF}$, $SU(5)$ and $SU(3)_{MF}$ - depending on which
components their VEVs are developed.

Further, there may be scalars which are consisted of three preons combined
with metaflavorless fermions $S_{L}^{[\boldsymbol{ab}]}$ and $S_{R}^{\prime
\lbrack \boldsymbol{a}^{\prime }\boldsymbol{b}^{\prime }]}$
\begin{eqnarray}
\Phi _{\lbrack ijk]}(x) &\propto &\epsilon _{\boldsymbol{abcde}}\left(
P_{iL}^{\boldsymbol{a}}CP_{jL}^{\boldsymbol{b}}\right) \left( P_{kL}^{%
\boldsymbol{c}}CS_{L}^{[\boldsymbol{de}]}\right) \text{ , \ }  \notag \\
\Phi _{\lbrack ijk]}^{\prime }(x) &\propto &\epsilon _{\boldsymbol{a}%
^{\prime }\boldsymbol{b}^{\prime }\boldsymbol{c}^{\prime }\boldsymbol{d}%
^{\prime }\boldsymbol{e}^{\prime }}\left( Q_{iR}^{\boldsymbol{a}^{\prime
}}CP_{jR}^{\boldsymbol{b}^{\prime }}\right) \left( Q_{kR}^{\boldsymbol{c}%
^{\prime }}CS_{R}^{\prime \lbrack \boldsymbol{d}^{\prime }\boldsymbol{e}%
^{\prime }]}\right)  \label{f3}
\end{eqnarray}%
which belong\ to the third-rank antisymmetric representation of the $%
SU(8)_{MF}$. Thereby, apart from preons, being the actual carriers of
metaflavors and metacolors, we have used a pair of elementary sterile
fermion multiplets, $S_{L}^{[\boldsymbol{ab}]}$ and $S_{R}^{\prime \lbrack
\boldsymbol{a}^{\prime }\boldsymbol{b}^{\prime }]}$ having only the $%
SO(5)_{MC}^{L}$ and $SO(5)_{MC}^{R}$ metacolors, respectively. We called
these fermions "sterilons" above in the section 2 treating them as possible
massless or light gauginos in the supersymmetrized metacolor theory\footnote{%
Remarkably, as one can notice from equations (\ref{wf}), (\ref{wf1}) and (%
\ref{f3}), our basic composites, both fermions $\Psi _{\lbrack jk]L}^{i}$ ($%
\Psi _{\lbrack jk]R}^{\prime i}$) and scalars $\Phi _{\lbrack ijk]}$ ($\Phi
_{\lbrack ijk]}^{\prime }$) contain the gauge superfield components of the
would-be supersymmetric metacolor theory - gauge vector fields $\mathbf{A}%
_{\mu }^{[\boldsymbol{ab}]}$ ($\mathbf{A}_{\mu }^{\prime \lbrack \boldsymbol{%
a}^{\prime }\boldsymbol{b}^{\prime }]}$) and gauginos $S_{L}^{[\boldsymbol{ab%
}]}$ ( $S_{R}^{\prime \lbrack \boldsymbol{a}^{\prime }\boldsymbol{b}^{\prime
}]}$), respectively.}. In principle, one might consider some other sterilon
containing scalar multiplets as well, particularly, the composite multiplets
$\Phi _{\lbrack jk]}^{i}(x)$ and $\Phi _{\lbrack jk]}^{\prime i}(x)$\
consisting of two preons and anti-preon, rather than three preons as in (\ref%
{f3}). However, it is clear that such scalars, if existed, would be
extremely unstable dissociating into our massless fermion multiplets $\Psi
_{\lbrack jk]L}^{i}$ and $\Psi _{\lbrack jk]R}^{\prime i}$ and sterilons
(screened by the corresponding metagluons). The same could be said about the
one-preon scalar multiplets $\Phi _{i}(x)$ and $\Phi _{i}^{\prime }(x)$
containing sterilons together with metagluons - they will readily decay into
the screened preons and sterilons. So, there are practically left only the
third-rank scalars (\ref{f3}) as some simple choice for the sterilon
containing scalar multiplets. The $SU(5)\times SU(3)_{MF}$ decomposition of
them gives%
\begin{equation}
56=(\overline{10},1)+(10,3)+(5,\overline{3})+(1,1)  \label{f33}
\end{equation}%
that means an immediate breaking of the starting symmetry $SU(8)_{MF}$ down
to $SU(5)\times SU(3)_{MF}$ when their VEVs are developed on the component $%
(1,1)$.

All the above basic scalars (\ref{s2}), (\ref{s3}) and \ref{f3}) develop
large VEVs which determine in general some effective mass scale in the
theory. They all provide an appropriate breaking of the $SU(8)_{MF}$ GUT
down to the Standard Model. Further breaking may be related to the exotic
scalars consisting of four preons
\begin{eqnarray}
\varphi _{\lbrack ijkl]}(x) &\propto &\left( P_{[iL}^{\boldsymbol{a}%
}CP_{jL}^{\boldsymbol{a}}\right) \left( P_{kL}^{\boldsymbol{b}}CP_{lL]}^{%
\boldsymbol{b}}\right) \text{ , \ }  \notag \\
\varphi _{\lbrack ijkl]}^{\prime }(x) &\propto &\left( Q_{[iR}^{\boldsymbol{a%
}^{\prime }}CQ_{jR}^{\boldsymbol{a}^{\prime }}\right) \left( Q_{kR}^{%
\boldsymbol{b}^{\prime }}CQ_{lR]}^{\boldsymbol{b}^{\prime }}\right)
\label{f4}
\end{eqnarray}%
where together with pure antisymmetrical multiplets (\ref{f4}) there could
also be many other ones with a mixed symmetry. They all are similar to the
exotic states of QCD, like the $(q\overline{q})(q\overline{q})$ states and
others. Even though they may still be bound, it is conceivable that the
significantly weaker four-preon attraction makes these very unstable scalars
to develop a hierarchically small VEV just needed for a conventional
breaking of the Standard Model\footnote{%
This implies in general some special fine tuning in total potentials of all
the scalar fields involved to provide, apart the small VEVs of the Higgs
multiplets (\ref{f4}), also the proper doublet-triplet splittings in them so
as to have the light electroweak doublets and heavy color triplets which
would mediate a proton decay. Interestingly, while the first problem is
generically solved in supersymmetric extension of any $SU(N)$\ grand
unification, a solution to the second one in the framework of the so-called
missing VEV conjecture \cite{dim} naturally appears only for the
supersymmetric $SU(8)$ GUT \cite{kob}.}. The antisymmetrical fourth-rank
multiplets (\ref{f4}) are in fact the self-conjugated multiplets of the $%
SU(8)_{MF}$ with the $SU(5)\times SU(3)_{MF}$ structure%
\begin{equation}
70=(\overline{5},1)+(5,1)+(10,\overline{3})+(\overline{10},3)  \label{f44}
\end{equation}%
containing the $SU(5)$ quintets which could break down the Standard Model
itself, and also give masses to quarks and leptons. So, using all the
effective scalar fields listed above one may finally come to the realistic
breaking pattern of the starting metaflavor symmetry $SU(8)_{MF}$.

\subsection{Gravitational conversion of vectorlike fermions}

In conclusion, let us note that the whole $L$-$R$ symmetric $SU(8)_{MF}$
metaflavor theory so far considered certainly appears pure vectorlike for
the identical $L$-preon and $R$-preon composite multiplets (\ref{216}) whose
masslessness is individually provided by their own chiral symmetry
preservation. Nonetheless, while preons themselves are left massless being
protected by their own metacolors, these left-handed and right-handed
composites being metacolor singlets may acquire masses. Indeed, they will
pair up with each other in (\ref{216}) and, thereby, acquire nonzero Dirac
masses. This presumably happens due to the gravitational transitions caused
by virtual black holes acting as intermediate particles, just as they
presumably do in gravitationally induced proton decay \cite{bh}. In fact,
the above transitions between the similar $L$-preon and $R$-preon composite
multiplets in (\ref{216}) can readily proceed via virtual black holes in a
way that all local symmetries involved are preserved. In fact, these
transitions basically depend on the compositeness scale $\Lambda _{MC}$ and
scale of quantum gravity determined, as usual, by the Planck mass $M_{Pl}$ $%
\simeq 1.2\cdot 10^{19}GeV$. As we find later, the scale $\Lambda _{MC}$
appears close to the Planck scale, due to which masses of all the composites
involved are expected to be very heavy. This would make our $SU(8)_{MF}$
metaflavor theory meaningless unless the proposed $L$-$R$ symmetry is broken
so as to allow some submultiplets in (\ref{216'}) to be left truly massless.
One could hope that such breaking may eventually exclude the right-handed
submultiplet $(\overline{5}+10,$ $\overline{3})_{R}$ in the composite
spectrum (\ref{216'}), while leaving there its left-handed counterpart, $(%
\overline{5}+10,$ $\overline{3})_{L}$, which can be then uniquely associated
with the observed three families of ordinary quarks and leptons. We consider
all that in detail in the next section.

\section{Composites - partially broken L-R symmetry}

\subsection{Strategy}

We propose a spontaneous breaking of the chiral symmetry $K(8)$ (\ref{881})
in the right-handed preon sector to induce a basic $L$-$R$ asymmetry in the
emerged $SU(8)_{MF}$ metaflavor theory being as yet vectorlike. We will
analyze this breaking first in the framework of the spectator gauge $%
SU(8)_{L}\times SU(8)_{R}$ model switching off the metaflavor interactions
and then consider its physical consequences in our $SU(8)_{MF}$ theory.
Actually, this breaking may appear due to a possible condensation of massive
composite scalars which unavoidably appear in the theory together with
composite fermions. We will see that just due to compositeness of the
scalars involved, while such breaking does not change the chiral symmetry of
the preons themselves, it may crucially change the chiral symmetry of their
composites. Some of these scalars (\ref{s2}), (\ref{s3}), (\ref{f3}) and (%
\ref{f4}) composed, respectively, from the left-handed ($P$) and
right-handed ($Q$) preons we considered in the previous section. The preon
metacolor interactions may generally produce at large distances (beyond a
confinement area) some effective potential due to which these scalars
develop VEVs violating global and local symmetries involved. We propose that
the third-rank scalars (\ref{f3}), $\Phi _{\lbrack ijk]}$ and $\Phi
_{\lbrack ijk]}^{\prime }$, containing the sterilons $S_{L}^{[\boldsymbol{ab}%
]}$ and $S_{R}^{\prime \lbrack \boldsymbol{a}^{\prime }\boldsymbol{b}%
^{\prime }]}$ may develop the largest VEVs\ in the theory that is a somewhat
natural breaking pattern in our $SO(5)_{MC}^{L}\times SO(5)_{MC}^{R}$
metacolor theory.

If from those two scalars only the $Q$-preon scalar multiplet $\Phi
_{\lbrack ijk]}^{\prime }$ develops VEV one comes, as we show below, to the
spontaneous violation of the $K(8)$ symmetry (\ref{881}) including its
center (\ref{881'}). This violation will presumably have a form
\begin{eqnarray}
K(8) &\rightarrow &[SU(8)]_{L}\times \lbrack SU(5)\times SU(3)]_{R}\text{ ,\
}  \notag \\
C(8,8) &\rightarrow &Z_{8}^{L}\times (Z_{5}\times Z_{3})^{R}\text{ }
\label{l-r}
\end{eqnarray}%
that breaks the chiral symmetry of right-handed $Q$-preon composites, while
the chiral symmetry of the left-handed $P$-preon composites is left intact.
This in turn means that the underlying $L$-$R$ symmetry becomes
spontaneously broken in the theory at large distances.

\subsection{How may it work}

The breaking pattern (\ref{l-r}) may be caused by some typical $L$-$R$
symmetric polynomial potential for scalars $\Phi _{\lbrack ijk]}$ and $\Phi
_{\lbrack ijk]}^{\prime }$

\begin{equation}
U=-M_{U}^{2}(\Phi ^{2}+\Phi ^{\prime 2})+h_{1}(\Phi ^{2}\Phi ^{2}+\Phi
^{\prime 2}\Phi ^{\prime 2})^{2}+h_{2}(\Phi ^{4}+\Phi ^{\prime
4})+h_{3}(\Phi ^{2}\Phi ^{\prime 2})+\cdot \cdot \cdot \text{ ,}  \label{u}
\end{equation}%
(with the notations $\Phi ^{2}=Tr(\Phi ^{+}\Phi )$, $\Phi ^{\prime
2}=Tr(\Phi ^{\prime +}\Phi ^{\prime })$, $\Phi ^{4}=Tr(\Phi ^{+}\Phi \Phi
^{+}\Phi )$ and $\Phi ^{\prime 4}=Tr(\Phi ^{\prime +}\Phi ^{\prime }\Phi
^{\prime +}\Phi ^{\prime })$ used) which presumably are induced by the
multi-preon interactions at large distances. Likewise, these interactions
may also produce the Yukawa preon-scalar potential of the type%
\begin{eqnarray}
L_{Y} &=&\frac{1}{M_{Y}^{3}}{\LARGE \{}\left[ \epsilon _{\boldsymbol{abcde}%
}\left( P_{iL}^{\boldsymbol{a}}CP_{jL}^{\boldsymbol{b}}\right) \left(
P_{kL}^{\boldsymbol{c}}CS_{L}^{[\boldsymbol{de}]}\right) \right] \Phi
^{\lbrack ijk]}  \notag \\
&&+\left[ \epsilon _{\boldsymbol{a}^{\prime }\boldsymbol{b}^{\prime }%
\boldsymbol{c}^{\prime }\boldsymbol{d}^{\prime }\boldsymbol{e}^{\prime
}}\left( Q_{iR}^{\boldsymbol{a}^{\prime }}CP_{jR}^{\boldsymbol{b}^{\prime
}}\right) \left( Q_{kR}^{\boldsymbol{c}^{\prime }}CS_{R}^{[\boldsymbol{d}%
^{\prime }\boldsymbol{e}^{\prime }]}\right) \right] \Phi ^{\prime \lbrack
ijk]}{\LARGE \}}+\cdot \cdot \cdot  \label{sup}
\end{eqnarray}%
being initially $L$-$R$ symmetric as well. These potentials including in
principle all possible high dimension couplings (denoted above by dots) are
evidently non-renormalizable and can be only considered as an effective
approach being valid at sufficiently low energies\footnote{%
These multi-preon interaction model looks somewhat similar to the well-known
multi-fermion interaction schemes used in the other contexts for chiral
symmetry breaking \cite{io} or spontaneous Lorentz violation \cite{jb}.}.
All dimensionful couplings with the mass parameters $M_{U}$, $M_{Y}$ and
others are proportional to appropriate powers of some UV cutoff which in our
case can be ultimately related to the preon confinement energy scale $%
\Lambda _{MC}$.

As in the known left-right models \cite{moh}, in the potential (\ref{u}) for
a natural range of the parameters, particularly, for $h_{3}>2(h_{1}+h_{2})$
and properly chosen the higher dimension coupling constants, the scalars $%
\Phi _{\lbrack ijk]}$ and $\Phi _{\lbrack ijk]}^{\prime }$ given in (\ref{f3}%
) and \ref{f33}) may develop the totally asymmetric VEV configuration%
\begin{equation}
\left\langle \Phi ^{\lbrack ijk]}\right\rangle =0\text{ , \ }\left\langle
\Phi ^{\prime \lbrack ijk]}\right\rangle =\delta _{p}^{i}\delta
_{q}^{j}\delta _{r}^{k}\epsilon ^{pqr}M_{LR}\text{ \ \ \ }(p,q,r=1,2,3)
\label{ph}
\end{equation}%
This VEV containing the antisymmetric third-rank tensor $\epsilon ^{pqr}$ of
the $SU(3)$ group certainly leads to the chiral symmetry breaking for the
right-handed composites, $SU(8)_{R}\rightarrow \lbrack SU(5)\times
SU(3)]_{R} $, leaving intact the $SU(8)_{L}$ symmetry for the left-handed
ones. In terms of the spectator gauge chiral symmetry, this means that all
non-diagonal gauge bosons related to the broken generators of the coset $%
SU(8)_{R}\boldsymbol{\ }/$ $[SU(5)\times SU(3)]_{R}$ acquire at large
distances the scale $M_{LR}$ order masses. That is, these bosons having been
strictly transversal at small distances acquire longitudinal parts
containing the composite Goldstone modes beyond the preon confinement area.
This means that, though the massless right-handed preons still possess the $%
SU(8)_{R}$ symmetry, the masslessness of their composites at large distances
is now solely controlled by its remained $[SU(5)\times SU(3)]_{R}$ part.
Indeed, in contrast to the conserved gauge currents of the preons, the gauge
currents of their composites related to the broken generators are no longer
conserved even in the classical Lagrangian. The more so, the triangle
anomalies of composites related to generators of the spectator gauge $%
SU(8)_{R}$ symmetry, other than those of its $[SU(5)\times SU(3)]_{R}$
subgroup, are no longer matched with anomalies of preons. In other words,
the spectator fermions which cancel anomalies of the right-handed preons
cannot cancel all anomalies of their composites. In fact, anomalies related
to the triangle graphs of composites with massive spectator gauge bosons
having longitudinal parts cannot be cancelled. This causes a proper
reduction of a possible chiral symmetry for right-handed preons and their
massless composites, just as is shown above in (\ref{l-r}), both
spontaneously and through the anomalies. Remarkably, the latter is crucially
related to the compositenes of the scalars $\Phi _{\lbrack ijk]}$ and $\Phi
_{\lbrack ijk]}^{\prime }$ which are solely emerged at large distances. For
elementary scalars existing at all distances one would only have a
spontaneous symmetry breaking of the chiral symmetry that could not change
the anomaly matching condition for right-handed preons and their composites.

A similar symmetry breaking effect related to the anomalies can be clearly
seen as well by considering the new four-preon interaction%
\begin{equation}
\frac{M_{LR}}{M_{Y}^{3}}\text{ }\epsilon ^{pqr}\epsilon _{\boldsymbol{a}%
^{\prime }\boldsymbol{b}^{\prime }\boldsymbol{c}^{\prime }\boldsymbol{d}%
^{\prime }\mathbf{e}^{\prime }}\left( Q_{pR}^{\boldsymbol{a}^{\prime
}}CQ_{qR}^{\boldsymbol{b}^{\prime }}\right) \left( Q_{rR}^{\boldsymbol{c}%
^{\prime }}CS_{R}^{\prime \lbrack \boldsymbol{d}^{\prime }\mathbf{e}^{\prime
}]}\right)  \label{n}
\end{equation}%
which follows from the Yukawa potential terms in (\ref{sup}) once the
asymmetric VEV (\ref{ph}) is developed ($Q_{pR}^{\boldsymbol{a}^{\prime }}$
are the right-handed family preon species, $p=1,2,3$). One can readily see
that the interaction (\ref{n}) possessing only $[SU(5)\times SU(3)]_{R}$
symmetry will necessarily modify the AM condition for right-handed states at
large distances. Indeed, this four-preon coupling induces radiative
corrections to the triangle graphs with circulation of composites containing
the family preons $Q_{pR}^{\boldsymbol{a}^{\prime }}$. As a result,
anomalies related to the graphs with the non-diagonal spectator gauge bosons
are not longer matched with those for the right-handed preon themselves.
There is left only the $[SU(5)\times SU(3)]_{R}$ symmetry correspondence
between short and large distances.

\subsection{Metaflavor theory with broken $L$-$R$ symmetry}

Let us now turn to the metaflavor interactions having been so far switched
off and see what happens in our $SU(8)_{MF}$ metaflavor gauge theory. One
can readily see that the asymmetric VEV configuration (\ref{ph}) related to
the third-rank scalars $\Phi _{\lbrack ijk]}$ and $\Phi _{\lbrack
ijk]}^{\prime }$ may lead to the natural spontaneous violation of the
starting $L$-$R$ symmetry in the $SU(8)_{MF}$ theory. First of all, this
symmetry itself becomes, as usual, spontaneously broken at the scale $M_{LR}$
down to the $SU(5)\times SU(3)$ symmetry group which we may identify as the
conventional $SU(5)$ grand unification times the $SU(3)_{F}$ family
symmetry. The point is, however, that the non-diagonal gauge bosons
receiving masses due to this symmetry breaking are no longer elementary at
large distances. Indeed, they acquire the longitudinal parts containing the
composite Goldstone modes which consist of the right-handed preons, as
follows from the VEV given above (\ref{ph}). Thereby, these modes will only
couple to the right-handed preon composites leaving the left-handed ones
basically intact. As a result, while the $SU(8)_{MF}^{3}$ triangle anomalies
related to $P$- and $Q$-preons automatically cancel each other, the same
anomalies related to the emerged left-handed and right-handed composites are
no longer cancelled. This means that the gauge $SU(8)_{MF}$ metaflavor
symmetry, similar to the chiral $SU(8)_{R}$ discussed above, not only
spontaneously breaks in an ordinary way, but also breaks by anomalies.
Thereby, while one has the $SU(8)_{MF}$ symmetry at small distances, there
only exist the $SU(5)\times SU(3)_{F}$ metaflavor symmetry beyond the preon
confinement area. This symmetry will then spontaneously breaks down to the
Standard Model by the other composite scalars (\ref{s2}), (\ref{s3}) and (%
\ref{f4}) introduced above, as is discussed in the next section. We propose
that, unlike the third-rank scalars $\Phi _{\lbrack ijk]}$ and $\Phi
_{\lbrack ijk]}^{\prime }$ (\ref{f3}) considered here, these scalars do not
break the $L$-$R$ symmetry. In fact, this may readily appear, if in their
potentials of the type used above (\ref{u}) the analogous parameters are
arranged in an opposite way. As a result, all such $P$-preon and $Q$-preon
scalars will develop the similar VEVs preserving the $L$-$R$ symmetry.

\subsection{Spectrum of survived composites}

Finally, let us consider the whole spectrum of composites to which the
spontaneous $L$-$R$ symmetry violation may lead in the survived $SU(5)\times
SU(3)_{F}$ metaflavor theory. As we have seen, while there still remains the
starting chiral symmetry $SU(8)_{L}$ for the left-handed preons and their
composites\footnote{%
Note that the effective chiral symmetry for the left-handed composites
(emerging in an absence of their metaflavor interactions) is in fact higher
than their actual metaflavor symmetry $SU(5)\times SU(3)_{F}$.}, for the
right-handed preon composite states\ we only have the broken symmetry given
in (\ref{l-r}). Therefore, whereas nothing changes for the $L$-preon
composites, the $R$-preons will only compose some particular submultiplets
in \ref{216'}) according to the matching condition for the broken chiral
symmetry (\ref{l-r}). In general, these submultiplets may not include the
three right-handed quark-lepton families $(\overline{5}+10,$ $\overline{3}%
)_{R}$. We might simply postulate it in the $L$-$R$ symmetry broken phase as
some possible ansatz being allowed by the different chiral symmetries of the
$L$-preon and $R$-preon composites. Nonetheless, one can try to derive it
using the proposed discrete preon number matching condition (\ref{yy}) at
least as some heuristic argument. Note that after the $L$-$R$ symmetry
breaking (\ref{ph}), together with the chiral symmetry $K(8)$ its discrete
center group, though being remained for preons, is properly reduced for
composites, just as is shown in (\ref{l-r}). Particularly, the discrete $%
Z_{8}^{R}$ symmetry of the right-handed composites comes to
\begin{equation}
Z_{8}^{R}\rightarrow Z_{5}^{R}\times Z_{3}^{R}  \label{pr}
\end{equation}%
while the $Z_{8L}$ symmetry of the left-handed composites is left intact.
Furthermore, in accordance with the matching condition for the broken chiral
symmetry $[SU(5)\times SU(3)]_{R}$ for R-preons and their composites, $%
Z_{5}^{R}$ and $Z_{3}^{R}$ can also be considered as discrete symmetries of
the right-handed quintet preons\ $Q_{sR}$ ($s=1,...,5$) of $SU(5)_{R}$ and
triplet preons $Q_{aR}$ ($a=1,2,3$) of $SU(3)_{R}$, respectively, which are
thereby separated. Namely, the $Q$-preon discrete symmetry in the broken $L$-%
$R$ symmetry phase is viewed from the large distances as the product (\ref%
{pr}) rather than the universal $Z_{8}^{R}$ symmetry for all eight preons,
as was in its unbroken phase (\ref{ds}). Now, if we require the discrete
preon charge matching (\ref{yy}) for preons and composites, the states
collected in the submultiplet $(\overline{5}+10,$ $\overline{3})_{R}$ will
never appear in physical spectrum since they possess both of discrete $%
Z_{5}^{R}$ and $Z_{3}^{R}$ charges, while the preons $Q_{sR}$ and $Q_{aR}$
have, by definition, only one of them. Indeed, if the $Z_{5}^{R}$ and $%
Z_{3}^{R}$ transformations are defined for $Q$-preons as
\begin{equation}
Q_{sR}\rightarrow e^{i2\pi q_{R}^{(5)}/5}Q_{sR}\text{ , \ }Q_{aR}\rightarrow
e^{i2\pi q_{R}^{(3)}/3}Q_{aR}  \label{sa}
\end{equation}%
then the $Z_{5}^{R}\times Z_{3}^{R}$ charges for them are $\left(
q_{R}^{(5)},0\right) $ and $\left( 0,q_{R}^{(3)}\right) $, while for their
composite submultiplets $(\overline{5},$ $\overline{3})_{R}$ and $(10,%
\overline{3})_{R}$ in (\ref{216'}) they come to $\left(
-q_{R}^{(5)},2q_{R}^{(3)}\right) $ and $\left(
2q_{R}^{(5)},-q_{R}^{(3)}\right) $, respectively. At the same time, all
other composite right-handed submultiplets in (\ref{216'}) readily match the
$Z_{5}^{R}\times Z_{3}^{R}$ charges for preons (normally, they may be taken
to be unit charges, $q_{R}^{(5)}=q_{R}^{(3)}=1$).

One way or another, the simplest combination of the right-handed
submultiplets in (\ref{216'}) which may simultaneously satisfy the AM
conditions (\ref{am2}) implied for the $[SU(5)\times SU(3)]_{R}$ symmetry,
as well as the above discrete preon number matching condition (\ref{yy}) is
in fact given by the collection
\begin{equation}
\lbrack
(45,1)_{R}+(5,8+1)_{R}+(1,3)_{R}]_{216}+2[(5,1)_{R}+(1,3)_{R}]_{8}+2(1,3)_{R}
\label{45}
\end{equation}%
The first and second terms (\ref{45}) contain submultiplets following from
the multiplets $216_{R}$ and $8_{R}$ in (\ref{216'}), respectively, while
the submultiplet $(1,3)_{R}$ has to appear two times more in order to
appropriately restore the anomaly coefficient balance for the $R$-preon
composites. The massless states (\ref{45}), though they are individually
protected by preservation of the chiral symmetry $[SU(5)\times SU(3)]_{R}$,
will generally pair up with the similar left-handed submultiplets in $%
216_{L}+8_{L}^{(1)}+8_{L}^{(2)}$ (\ref{216'})%
\begin{eqnarray}
&&\left[ (\overline{5}+10,\text{ }\overline{3}%
)_{L}+(45,1)_{L}+(5,8+1)_{L}+(24,3)_{L}+(1,3)_{L}+(1,\overline{6})_{L}\right]
_{216}  \notag \\
&&+\text{ }2\left[ (5,1)_{L}+(1,3)_{L}\right] _{8}  \label{45'}
\end{eqnarray}%
due to the quantum gravitational conversion discussed in the section 4.4. As
a result, due to the expected closeness of the compositeness scale $\Lambda
_{MC}$ to the Planck scale $M_{Pl}$, the Dirac masses of all such composites
appear very heavy. Eventually, only some special submultiplets in the
starting left-handed and right-handed composite multiplets (\ref{216'})
including those which contain ordinary quarks and leptons may emerge at
present laboratory energies.

\section{Some immediate physical consequences}

\subsection{Quarks, leptons and beyond}

We have seen in the section 5.2 above that the $L$-$R$ symmetry breaking in
the right-handed preon sector initiated by the composite scalar condensation
(\ref{ph}) makes the starting metaflavor symmetry $SU(8)_{MF}$ to be reduced
at large distances to the product of the standard $SU(5)$ GUT and family
symmetry $SU(3)_{F}$
\begin{equation}
SU(8)_{MF}\rightarrow SU(5)\times SU(3)_{F}\text{ }  \label{gf}
\end{equation}
This $SU(5)\times SU(3)_{F}$ symmetry of composites appearing in the $L$-$R$
symmetry broken phase is in essence the chiral remnant of the initial
vectorlike $SU(8)_{MF}$ symmetry of preons that only exists at small
distances and breaks by anomalies at the large ones. Accordingly, the
massless composite fermions survived during the gravitational conversion of
similar $L$-preon and $R$-preon composites in (\ref{45}, \ref{45'}) and
decoupling them from a low-energy spectrum are given now by the collection
of the $SU(5)\times SU(3)_{F}$ multiplets
\begin{equation}
(\overline{5}+10,\overline{3})_{L}+(24,3)_{L}+(1,\overline{6})_{L}+2(1,3)_{R}
\label{fm}
\end{equation}%
which automatically appear free from both the $SU(5)$ and $SU(3)_{F}$
anomalies.

The first term in (\ref{fm}) contains just three conventional families of
quarks and leptons which are presented below
\begin{equation}
\begin{array}{ccc}
\left(
\begin{array}{c}
u \\
d%
\end{array}%
\right) _{L}\text{ \ \ \ \ \ }\left(
\begin{array}{c}
c \\
s%
\end{array}%
\right) _{L}\text{ \ \ \ \ \ }\left(
\begin{array}{c}
t \\
b%
\end{array}%
\right) _{L} & {\small \longleftrightarrow } & {\small \mathcal{C}}_{l}%
{\small \mathcal{W}}_{a}\overline{\mathcal{F}}^{p} \\
&  &  \\
\widetilde{u}_{L}\text{ \ \ \ \ \ \ \ \ \ \ \ \ \ \ }\widetilde{c}_{L}\text{
\ \ \ \ \ \ \ \ \ \ \ \ \ \ \ }\widetilde{t}_{L} & {\small %
\longleftrightarrow } & {\small \epsilon }^{lmn}{\small \mathcal{C}}_{m}%
{\small \mathcal{C}}_{n}\overline{\mathcal{F}}^{p} \\
&  &  \\
\widetilde{d}_{L}\text{ \ \ \ \ \ \ \ \ \ \ \ \ \ \ }\widetilde{s}_{L}\text{
\ \ \ \ \ \ \ \ \ \ \ \ \ \ \ \ }\widetilde{b}_{L} & {\small %
\longleftrightarrow } & \overline{{\small \mathcal{C}}}^{l}{\small \mathcal{F%
}}_{p}{\small \mathcal{F}}_{q}{\small \epsilon }^{pqr} \\
&  &  \\
&  &  \\
\left(
\begin{array}{c}
e \\
-\nu _{e}%
\end{array}%
\right) _{L}\text{\ \ \ \ \ }\left(
\begin{array}{c}
\mu \\
-\nu _{\mu }%
\end{array}%
\right) _{L}\text{\ \ \ \ \ }\left(
\begin{array}{c}
\tau \\
-\nu _{\tau }%
\end{array}%
\right) _{L} & {\small \longleftrightarrow } & \overline{\mathcal{W}}^{a}%
{\small \mathcal{F}}_{p}{\small \mathcal{F}}_{q}{\small \epsilon }^{pqr} \\
&  &  \\
\widetilde{e}_{L}\text{ \ \ \ \ \ \ \ \ \ \ \ \ }\widetilde{\mu }_{L}\text{
\ \ \ \ \ \ \ \ \ \ \ \ \ }\widetilde{\tau }_{L} & {\small %
\longleftrightarrow } & {\small \varepsilon }^{ab}{\small \mathcal{W}}_{a}%
{\small \mathcal{W}}_{b}\overline{\mathcal{F}}^{p}%
\end{array}%
\bigskip  \label{qp}
\end{equation}%
On the right side one can see their preon content in terms of the
left-handed Weil spinor fields $\mathcal{C}_{l}$, $\mathcal{W}_{a}$ and $%
\mathcal{F}_{p}$ with a proper implication of the antisymmetric tensors $%
\epsilon ^{lmn}$, $\epsilon ^{pqr}$, $\varepsilon ^{ab}$ of the groups $%
SU(3)_{C}$, $SU(3)_{F}$ and $SU(2)_{W}$, respectively. Here, $\mathcal{C}%
_{l} $ ($l,m,n$ $=1,2,3$) stand for the "chromons" being the basic triplet
of the color $SU(3)_{C}$, then $\mathcal{W}_{a}$ ($a,b=1,2$) denotes the
"isons" forming the elementary doublet of the weak isotopic spin $SU(2)_{W}$
and, finally, the family number carriers are given by the "famons" $\mathcal{%
F}_{p}$ ($p,q,r=1,2,3$) which belong to the fundamental triplet of the
family symmetry $SU(3)_{F}$. Note that in accordance with a usual assignment
of quarks and leptons to the left-handed multiplets $(\overline{5}+10)_{L}$
in the $SU(5)$ GUT, the right-handed quark and lepton states are taken in (%
\ref{qp}) in terms of their anti-states taken in the left-handed basis.
Also, the lepton doublets are presented in terms of the anti-doublets, as
they are usually treated in the anti-quintet $\overline{5}_{L}$ of the $%
SU(5) $. From the table (\ref{qp}) one can readily find the electric charges
of chromons, isons and famons, respectively,
\begin{equation}
Q_{\mathcal{C}}=-1/3;\ \ Q_{\mathcal{W}_{1}}=1,\ Q_{\mathcal{W}_{2}}=0;\text{
\ }Q_{\mathcal{F}}=0.  \label{qqqq}
\end{equation}

It is important to note that the compositeness scale $\Lambda _{MC}$ for
universal preons composing both quarks and leptons appears too high to
directly observe their composite nature \cite{ans}. Indeed, one can readily
see from the quark-lepton preon structure shown above in (\ref{qp}) that the
quark pair $u+d$ contains the same preons as the antiquark-antilepton pair $%
\overline{u}+e^{+}$. This eventually would lead to the process%
\begin{equation}
u+d\rightarrow \overline{u}+e^{+}  \label{pd}
\end{equation}%
and consequently to the proton decay $p\rightarrow \pi ^{0}+e^{+}$ just due
to a simple rearrangement of preons inside the proton. To prevent this, the
compositeness scale $\Lambda _{MC}$ has to be of the order of the scale of
the $SU(5)$\ GUT or even larger, $\Lambda _{MC}\gtrsim M_{GUT}\approx 2\cdot
10^{16}$ $GeV$ \footnote{%
Everywhere in the paper we consider just the supersymmetric $SU(5)$\ GUT
scale, since the non-supersymmetric $SU(5)$ model is in fact ruled out by
observations \cite{ne}.}. This scale is also in accordance with the energy
scale at the end of inflation in a conventional cosmological scenario \cite%
{ne} due to which any potentially stable heavy composite will be safely
inflated out (like the magnetic monopoles involved in the theory). On the
other hand, most of such heavy composites can decay into ordinary quarks and
leptons by the virtual black hole mechanism described in the section 4.4
provided that all accompanying local symmetries are preserved.

\subsection{Quark-lepton masses and mixings}

The quarks and leptons presented above in the submultiplet $(\overline{5}+10,%
\overline{3})_{L}$ receive their masses from the $SU(8)_{MF}$ invariant
Yukawa couplings with composite scalars introduced in the section 4.3. As
was already mentioned above, in contrast to the third-rank scalars (\ref{f3}%
) causing the basic $L$-$R$ symmetry breaking in the theory, the other
composite scalars (\ref{s2}), (\ref{s3}) and \ref{f4}) develop presumably
pure $L$-$R$ symmetrical VEVs. Actually, in the $SU(8)_{MF}$ theory there is
some doubling of the identical scalar multiplets which are composed
individually from the left-handed and right-handed preons. We call them the $%
L$- and $R$-scalars to show that they selectively interact with the
left-handed and right-handed composite fermions, respectively. As matter of
fact, in the Yukawa couplings of the left-handed composite fermions
presented in (\ref{fm}) the $L$-scalars may only contribute. Particularly,
for the submultiplet $(\overline{5}+10,\overline{3})_{L}$ one has, as in a
conventional $SU(5)$ GUT, the two independent $SU(8)_{MF}$ invariant
couplings
\begin{eqnarray}
&&\frac{1}{\mathcal{M}}\left[ \Psi _{\lbrack jk]L}^{i}C\Psi _{\lbrack
mp]L}^{l}\right] \varphi ^{\lbrack jkmp]}(a_{u}\chi _{\lbrack il]}+b_{u}\chi
_{\{il\}})  \notag \\
&&\frac{1}{\mathcal{M}}\left[ \Psi _{\lbrack jk]L}^{i}C\Psi _{\lbrack
im]L}^{l}\right] \varphi ^{\lbrack jkmp]}(a_{d}\chi _{\lbrack lp]}+b_{d}\chi
_{\{lp\}})\text{ }  \label{udd}
\end{eqnarray}%
with different index contraction for the up quarks, and down quarks and
leptons, respectively. The mass $\mathcal{M}$ stands for some effective
scale in the theory that is related to the metacolor scale $\Lambda _{MC}$,
while $a_{u,d}$ and $b_{u,d}$ are some dimensionless constants of the order
of $1$. Note that these couplings contain two types of scalars with the
appropriate $SU(5)\times SU(3)_{F}$ components: the exotic four-preon $%
\varphi $ multiplet given in (\ref{f4}) and (\ref{f44}) containing the $%
SU(5) $ quintets $(5,1)_{u}$ and $(\overline{5},1)_{d}$ to break the
Standard Model at the electroweak scale $M_{SM}$\footnote{%
Therefore, in contrast to a conventional $SU(5)$ GUT our model contain the
two different scalar quintets stemming from the same self-conjugated
multiplet $\varphi ^{\lbrack ijkl]}$ of the $SU(8)_{MF}$. We denoted these
quintets as $(5,1)_{u}$ and $(\overline{5},1)_{d}$ to stress that they give
masses to the up and down quarks, respectively (see below).} and the basic
two-preon $\chi _{\lbrack ij]}$ and $\chi _{\{ij\}}$ multiplets in (\ref{s2}%
) and (\ref{s22}) with components $(1,\overline{3})$ and $(1,6)$ to properly
break the $SU(3)_{F}$ family symmetry at the large scale $M_{F}$. We call
them the "vertical" and "horizontal" scalars, respectively. Taken together,
they presumably determine masses and mixings of all quarks and leptons.
Otherwise, we would have to introduce the multi-preon scalar composites with
both vertical and horizontal indices that seems to be too complicated%
\footnote{%
In this case, there also would appear the stronger gauge hierarchy problem
to arrange the VEVs of all these scalars in an appropriate way.}. After the $%
SU(8)_{MF}$ symmetry breaking (\ref{gf}) the Yukawa couplings acquire the
transparent $SU(5)\times SU(3)_{F}$ invariant form (all metaflavor indices
are omitted)%
\begin{eqnarray}
&&\left[ (10,\overline{3})_{L}C(10,\overline{3})_{L}\right]
(5,1)_{u}[a_{u}(1,\overline{3})+b_{u}(1,6)]/\mathcal{M}\text{ ,}  \notag \\
&&[(\overline{5},\overline{3})_{L}C(10,\overline{3})_{L}](\overline{5}%
,1)_{d}[a_{d}(1,\overline{3})+b_{d}(1,6)]/\mathcal{M}\text{ }  \label{ud'}
\end{eqnarray}%
where we only include those components of the vertical scalar $\varphi $ and
horizontal scalars $\chi _{\lbrack ij]}$ and $\chi _{\{ij\}}$ which develop
the VEVs. Just the horizontal scalar VEVs determine through the Yukawa
couplings (\ref{ud'}) the mass matrices for quarks and leptons
\begin{eqnarray}
\widehat{m}_{pq}^{u} &=&\left\langle 5,1\right\rangle _{u}[a_{u}\left\langle
1,\overline{3}\right\rangle _{[pq]}+b_{u}\left\langle 1,6\right\rangle
_{\{pq\}}]/\mathcal{M}\text{ \ }  \notag \\
\widehat{m}_{pq}^{d} &=&\left\langle \overline{5},1\right\rangle
_{d}[a_{d}\left\langle 1,\overline{3}\right\rangle _{[pq]}+b_{d}\left\langle
1,6\right\rangle _{\{pq\}}]/\mathcal{M}\text{ \ }  \label{mn}
\end{eqnarray}%
where the angle brackets denote the corresponding VEVs, while $p,q=1,2,3$
stand for family indices. The matrices $\widehat{m}_{pq}^{u}$ and $\widehat{m%
}_{pq}^{d}$ are defined at the grand unified scale and have to be then
extrapolated down to an actual mass range for quarks and leptons\footnote{%
Note that hand in hand with a good relation between masses of the $b$-quark
and $\tau $-lepton there also appear in the matrix $\widehat{m}^{d}$, as
usual in the $SU(5)$ GUT, the bad relations for the lighter down quarks and
leptons. However, this readily can be alleviated in our model by introducing
in (\ref{udd}) an extra higher dimension Yukawa coupling which includes the
adjoint composite scalar $\phi _{i}^{j}$ (\ref{s3}) as well%
\begin{equation*}
\frac{1}{\mathcal{M}^{2}}\left[ \Psi _{\lbrack jk]L}^{i}C\Psi _{\lbrack
nm]L}^{l}\right] \phi _{i}^{n}\varphi ^{\lbrack jkmp]}(a_{d}^{\prime }\chi
_{\lbrack lp]}+b_{d}^{\prime }\chi _{\{lp\}})
\end{equation*}%
This will decouple masses of the first two quark-lepton families from each
other leaving intact masses of the third one.}. Depending on which
components the above symmetrical and asymmetrical VEVs are developed, one
comes to different texture zero types for all matrices involved. The strong
hierarchies of the quark-lepton masses and mixings may be now explained by
somewhat softer hierarchies between the breaking directions of the $%
SU(3)_{F} $ family symmetry whose scale $M_{F}$\ is imposed to be close to
the effective scale $\mathcal{M}$ (see some significant references in the
section 7).

\subsection{Neutrino masses}

While quarks and charged leptons acquire masses in a way described above,
the neutrinos are still left massless. However, in contrast to the standard $%
SU(5)$\ GUT our model contains some generic candidates for singlet heavy
neutrinos to eventually generate masses of the physical ones. Indeed, the
initially massless extra submultiplets $(24,3)_{L}$, $(1,\overline{6})_{L}$
and $(1,3)_{R}^{(1,2)}$ in the survived collection (\ref{fm}) become then
heavy acquiring the family scale order masses once the $SU(3)_{F}$ symmetry
breaks down. In order to sufficiently suppress all flavor-changing
transitions, which would be mediated by the family gauge bosons, this scale $%
M_{F}$ has to be at least of the order $10^{5}$ $GeV$, though in our model,
due to the VEVs developed by composite scalars, it may be as large as the
scale $M_{GUT}$ and even larger. The above submultiplets receive the heavy
Majorana masses from the $SU(8)_{MF}$ invariant Yukawa couplings%
\begin{equation}
\left[ \Psi _{\lbrack jk]L}^{i}C\Psi _{\lbrack im]L}^{j}\right] \left(
a_{ex}\chi ^{\lbrack km]}+b_{ex}\chi ^{\{km\}}\right) \text{ }  \label{rn}
\end{equation}%
for the left-handed composites and in a similar way for the right-handed
ones (the coupling constants $a_{ex}$ and $b_{ex}$ are dimensionless and of
the order of $1$). In the decomposed $SU(5)\times SU(3)_{F}$ invariant form
these couplings look as%
\begin{equation}
\lbrack (24,3)_{L}C(24,3)_{L}+(1,\overline{6})_{L}C(1,\overline{6}%
)_{L}][a_{ex}(1,3)+b_{ex}(1,\overline{6})]  \label{rn1}
\end{equation}%
when either the $SU(5)$ indices or the $SU(3)_{F}$ indices are only
contracted. As can be readily seen by comparing the Yukawa couplings (\ref%
{rn}) and (\ref{rn1}) with (\ref{udd}) and (\ref{ud'}), all extra
submultiplets in (\ref{fm}), acquire eventually the mass-matrices being
basically similar in form to those for quarks and leptons. Accordingly, some
significant mass hierarchies (by two or three orders of magnitude) between
extra state families are also expected.

Remarkably, the three SM\ singlet states in the submultiplet $(24,3)_{L}$
appearing in its $SU(3)_{C}\times SU(2)_{W}\times SU(3)_{F}$ decomposition
as $(1,1,3)_{L}$ could be considered as candidates for the heavy
right-handed neutrinos $\mathcal{N}_{R}^{p}$ taken as the left-handed
anti-neutrinos $\mathcal{N}_{Lp}^{c}$ in the considered basis ($p$ stands
for the family symmetry index, $p=1,2,3$). As mentioned above, they develop
masses%
\begin{equation}
m_{\mathcal{N}_{p}}=f^{p}M_{F}\text{, }f^{p}=(0.001-1)  \label{hi}
\end{equation}%
including a possible mass hierarchy between heavy neutrinos of different
families. The massless physical neutrinos contained in the submultiplet $(%
\overline{5},\overline{3})_{L}$ in (\ref{fm}) may mix with them, thus
acquiring the tiny Majorana masses through the see-saw mechanism \cite{moh}
in their common $6\times 6$ mass-matrix. Actually, this mixing may appear
from the high-dimensional $SU(8)_{MF}$ Yukawa couplings
\begin{equation}
\frac{1}{\mathcal{M}^{3}}\left[ \Psi _{\lbrack jk]L}^{i}C\Psi _{\lbrack
im]L}^{l}\right] \varphi _{\lbrack lpqr]}\left( \chi ^{\lbrack jp]}\chi
^{\lbrack kq]}\chi ^{\lbrack mr]}+\cdot \cdot \cdot \right)  \label{mx}
\end{equation}%
with the above vertical and horizontal scalars $\varphi $ and $\chi $
developing VEVs on the $SU(5)\times SU(3)_{F}$ components $(5,1)_{d}$ and $%
(1,3)$ plus $(1,\overline{6})$ , respectively (the dimensionful coupling
constant is properly determined here by the effective scale $\mathcal{M}$,
while dots stand for other terms with anti-symmetrical and symmetrical
horizontal scalars with all possible index contractions). This coupling
after electroweak and family symmetry breakings at the corresponding scales $%
M_{SM}$ and $M_{F}$ leads to the mixing terms for the submultiplets $(%
\overline{5},\overline{3})_{L}$ and $(24,3)_{L}$ being of the order
\begin{equation}
\left( M_{F}/\mathcal{M}\right) ^{3}M_{SM}\text{ }  \label{2}
\end{equation}%
that in turn induces masses for physical neutrinos. These masses according
to (\ref{hi}) and (\ref{2}) come to
\begin{equation}
m_{\nu _{p}}\sim \left( M_{F}/\mathcal{M}\right) ^{6}M_{SM}^{2}/f^{p}M_{F}
\label{mnu}
\end{equation}%
which appear to be of the right order provided that the scale of family
symmetry $M_{F}$ is of the order of the scale $\mathcal{M}$ that, as we have
seen above, is also required for masses of quarks and leptons. Apart from
that, the scale $M_{F}$ has to be close to the scale of the $SU(5)$\ GUT
given above to be in agreement with observed limitations on mass spectrum of
physical neutrinos \cite{ne}. One can see that since masses of heavy
neutrinos $\mathcal{N}_{p}$ follow in (\ref{rn}) the mass hierarchy of
quark-lepton families, masses of physical neutrinos have to obey the
inverted hierarchy so that the electron neutrino appears to be the heaviest
one with mass value up to $1eV$ or even larger.

\subsection{Heavy states}

The compositeness scale $\Lambda _{MC}$\ may on its own cause a limit on on
the composite fermion masses appearing as a result of the quantum
gravitational transitions between similar states in the right-handed
multiplets (\ref{45}) and left-handed multiplets (\ref{45'})
\begin{equation}
\lbrack
(45,1)_{L,R}+(5,8+1)_{L,R}+(1,3)_{L,R}]_{216}+2[(5,1)_{L,R}+(1,3)_{L,R}]_{8}
\label{fr}
\end{equation}%
into each other, as was argued in the section 4.4. From dimensional
arguments related to a general structure of the three-preon composites
proposed above in (\ref{wf}) and (\ref{wf1}), masses of these vectorlike
multiplets could be of the order
\begin{equation}
M_{V}\sim (\Lambda _{MC}/M_{Pl})^{9}\Lambda _{MC}  \label{hm}
\end{equation}%
that corresponds to the high-dimension interaction between left-handed and
right-handed composites of the type (the metacolor indices are omitted)%
\begin{equation}
\frac{1}{M_{Pl}^{9}}\left[ \left( \overline{P}_{R}^{j}\gamma _{\mu
}P_{iR}\right) \overline{P}_{R}^{k}F^{\mu \nu }\right] \left[ \left(
\overline{Q}_{L}^{i}\gamma ^{\rho }Q_{jL}\right) Q_{kL}F_{\rho \nu }^{\prime
}\right] +h.c.  \label{hd}
\end{equation}%
Similarly, the screened one-preon states
\begin{equation}
(5,1)_{L,R}^{(1,2)}+(1,3)_{L,R}^{(1,2)}  \label{fr2}
\end{equation}%
in (\ref{fr}) acquire masses when being pairing with each other through the
analogous high-dimension interactions of the $P$- and $Q$-preons including
their metagluons
\begin{equation}
\frac{1}{M_{Pl}^{7}}\left[ (F_{\mu \nu }F^{\mu \nu })\overline{P}_{L}^{i}%
\right] \left[ (F_{\rho \sigma }^{\prime }F_{\mu \nu }^{\prime \rho \sigma
})Q_{iR}\right] +h.c.  \label{hd1}
\end{equation}%
appearing from their structure given in (\ref{wf}) and (\ref{wf1}). Again,
from the dimensional arguments one may conclude that these masses has a
natural order%
\begin{equation}
(\Lambda _{MC}/M_{Pl})^{7}\Lambda _{MC}  \label{hm'}
\end{equation}%
that may be significantly larger than masses (\ref{hm}) of the 3-preon
states. These gravitationally induced masses are indeed very sensitive to
the confinement\ scale $\Lambda _{MC}$. In order for the vectorlike
multiplets (\ref{fr}) and (\ref{fr2}) to be enough heavy, say with masses $%
10^{5}GeV$ and higher, the scale $\Lambda _{MC}$ has to be close to the
Planck scale $M_{Pl}$, namely $\Lambda _{MC}\gtrsim 5\cdot 10^{17}GeV$, as
is readily seen from (\ref{hm}).

Note that some of heavy vectorlike submultiplets, particularly, $(5,8)_{L,R}$
in (\ref{fr}) may mix with the physical submultiplet $(\overline{5},%
\overline{3})_{L}$ in (\ref{fm}) which contains the lepton doublet and down
antiquarks. This mixing appears through the above Yukawa couplings (\ref{rn}%
) with simultaneous contractions the $SU(5)$ and $SU(3)_{F}$\ indices, that
leads instead of (\ref{rn1}) to the couplings
\begin{equation}
\lbrack (5,8)_{L}C(\overline{5},\overline{3})_{L}][a_{ex}(1,3)+b_{ex}(1,%
\overline{6})]  \label{rn2}
\end{equation}%
These couplings after the $SU(3)_{F}$ symmetry breaking cause the family
scale $M_{F}$ order\ mixing masses in the corresponding triangular mass
matrices (describing either left-right or right-left fermion component
transitions but not both simultaneously)\footnote{%
For simplicity, we do not consider analogous mixings of the physical
multiplet $(\overline{5},\overline{3})_{L}$ with the three-preon multiplets $%
(5,1)_{L,R}$ and screened one-preon preon multiplets $(5,1)_{L,R}^{(1,2)}$
in (\ref{fr}).}. Though these matrices cannot significantly disturb the
masses of all the states involved, they may induce a large admixture of the
heavy states in the left-handed leptons in $(\overline{5},\overline{3})_{L}$
that in turn could strongly influence their CKM matrix. To properly suppress
this mixing one has to generally propose that the non-diagonal mass term of
the heavy multiplets $(5,8)_{L,R}$ in (\ref{rn2}) is at least less than its
diagonal mass $M_{V}$ (\ref{hm}) induced by gravity, namely,
\begin{equation}
M_{F}\lesssim (\Lambda _{MC}/M_{Pl})^{9}\Lambda _{MC}  \label{mm}
\end{equation}%
For the naturally high family scale $M_{F}$ being at least of the order of
the $SU(5)$\ GUT scale, as is required from the masses of neutrinos
discussed above, the metacolor scale $\Lambda _{MC}$ in (\ref{mm}) has to
appear very close to the Planck scale $M_{Pl}$. Particularly, for $M_{F}\sim
M_{GUT}$ and $\Lambda _{MC}\sim M_{Pl}$ one has for an angle of the above
extra mixing
\begin{equation}
\tan \theta \sim M_{F}/M_{V}\sim (M_{F}/M_{Pl})(M_{Pl}/\Lambda _{MC})^{10}
\label{v}
\end{equation}%
a reasonably small value $\theta \sim M_{GUT}/M_{Pl}\sim 10^{-3}$.
Generally, depending on actual values of the scales $\Lambda _{MC}$ and $%
M_{F}$, there could be expected some marked violation of unitarity in the
CKM matrix for leptons which may be of a special interest for observations.
In contrast, for quarks, due to mixing of only right-handed down quarks in (%
\ref{rn2}), the CKM matrix is left practically intact.

\subsection{Basic scenarios}

Actually, there are two phases in the model: the $L$-$R$ symmetry phase with
the initial vectorlike $SU(8)_{MF}$ symmetry of the left-handed and
right-handed preons and the broken the $L$-$R$ symmetry phase with the $%
SU(5)\times SU(3)_{F}$ symmetry of the chiral composites having principally
different multiplet spectra for the right handed and left-handed states
given above in (\ref{45}, \ref{45'}). Accordingly, the massless composite
fermions survived after the gravitational conversion of similar $L$-preon
and $R$-preon composites are finally collected in the $SU(5)\times SU(3)_{F}$
multiplets (\ref{fm}). In contrast to preons, there is no the $SU(8)_{MF}$
unification for composites, because it is broken by anomalies at the large
distances, as we discussed in the section 5.

According to all the symmetry breakings involved, our preon model contains
in fact a few basic mass scales determined by the generic compositeness
scale $\Lambda _{MC}$. They are: the $L$-$R$ symmetry breaking scale $M_{LR}$
where the starting metaflavor symmetry $SU(8)_{MF}$ of preons breaks down to
the $SU(5)\times SU(3)_{F}$ symmetry for composites; then the grand
unification scale $M_{GUT}$ and the family scale $M_{F}$ where the $SU(5)$
and $SU(3)_{F}$ get broken, respectively. These breakings are provided by
the basic scalar composites (\ref{s2}), (\ref{s3}) and (\ref{f3}) developing
the large VEVs in the theory that eventually leads to the Standard Model.
Further breaking may be only related to the exotic scalars consisting of
four preons (\ref{f4}) which, as we proposed in the section 4.3, develop the
VEV of the electroweak scale order $M_{SM}$. The model predicts three types
of the composite fermion states which are: (1) the three families of
ordinary quarks and leptons $(\overline{5}+10,\overline{3})_{L}$ in (\ref{fm}%
) with masses at the electroweak scale $M_{SM}$, (2) the chiral multiplets $%
(24,3)_{L}+(1,\overline{6})_{L}+2(1,3)_{R}$ with the Majorana type masses at
the family scale $M_{F}$ and (3) the vectorlike multiplets (\ref{fr}) with
the gavitationally induced Dirac masses (\ref{hm}) and (\ref{hm'}). The
latter gives one more high scale $M_{V}$ in the theory which may be located
in a wide range of masses up to the Planck scale $M_{Pl}$. Further details
depend on a taken scenario with a particular interplay between the high mass
scales mentioned above.

The most natural scenario seems to be the case with the Planck scale as the
effective theory scale, $\mathcal{M}$ $\sim $ $M_{Pl}$, when all the high
scales are of the same order determined by the generic scale $\Lambda _{MC}$
which itself is taken near the Planck scale%
\begin{equation}
M_{LR}\sim M_{GUT}\sim M_{F}\sim M_{V}\sim \Lambda _{MC}\sim M_{Pl}
\label{sc1}
\end{equation}%
As a result, the starting $SU(8)_{MF}$ symmetry breaks at once to the
Standard Model at the Planck scale. This means that only ordinary quarks and
leptons will contribute to gauge coupling unification in the $SU(5)$ GUT
whose scale $M_{GUT}$ has to be increased now to the Planck mass $M_{Pl}$.
This, as usual, can be done by an embedding the whole model into $N=1$
supergravity \cite{moh} with a proper extension of particle spectra
involved. Due to the highest scales used, this scenario at low energies is
not much different from a conventional supergravity grand unification.
Actually, it additionally predicts only a strong mixing of lepton doublets
with heavy states given in (\ref{v}) with near maximal angle ($\theta \sim
\pi /4$) and, thereby, some significant violation of unitarity in the lepton
CKM matrix. Neither physically interesting neutrino masses, nor any extra
composites at intermediate scales are provided.

The next may be the case with the $SU(5)$ GUT scale as an effective theory
scale, $\mathcal{M}$ $\sim M_{GUT}$, where the $M_{GUT}$ is considered as a
conventional supersymmetric $SU(5)$ GUT scale. This is somewhat lower than
the compositeness scale $\Lambda _{MC}$ which cannot be less than $5\cdot
10^{17}GeV$, as was stated above. At the same time, the $L$-$R$ symmetry
breaking scale $M_{LR}$ (which at the same time is the scale of the $%
SU(8)_{MF}$ as well) should always be near the scale $\Lambda _{MC}$ to
cause the principally different composite spectra, (\ref{45}) and (\ref{45'}%
), for the left-handed and right-handed preons, respectively. If otherwise $%
M_{LR}<<\Lambda _{MC}$, only the similar multiplets $%
216_{[jk]L}^{i}+8_{iL}^{(1)}+8_{iL}^{(2)}$ and \ $%
216_{[jk]R}^{i}+8_{iR}^{(1)}+8_{iR}^{(2)}$ (\ref{216}) of the yet unbroken $%
SU(8)_{MF}$ metaflavor symmetry would appear in a composite spectrum. They
all then pair up with each other and, thereby, acquire heavy Dirac masses
that would make the whole model to be meaningless. So, we propose that only
the $L$-$R$ symmetry breaking scale $M_{LR}$ is of the Planck scale order,
others are less%
\begin{equation}
M_{LR}\sim \Lambda _{MC}\sim M_{Pl}\text{ , \ }M_{GUT}\sim M_{F}\sim M_{V}
\label{sc2}
\end{equation}%
that implies some two-three order of magnitude hierarchy between these two
groups of scales. In this scenario, one may have the right order masses for
physical neutrinos, as one can see from (\ref{mnu}). Also there is a
significant mixing of lepton doublets with heavy states mentioned above in (%
\ref{v}) provided that masses $M_{V}$ of the vectorlike multiplets is also
of the order of $M_{GUT}$. For that, the compositeness scale has to be $%
\Lambda _{MC}\approx 6\cdot 10^{18}$ GeV, as follows from (\ref{hm}). Thus,
the whole picture may look as follows. There are only $L$-preons and $R$%
-preons at the Planck scale with the metaflavor symmetry $SU(8)_{MF}$ which
runs over a short distance down to the compositeness scale $\Lambda _{MC}$
located a bit lower. Then at the scale $M_{LR}\sim \Lambda _{MC}$ this
symmetry breaks to the $SU(5)\times SU(3)_{F}$ in the $L$-$R$ asymmetric way
so that there appears two different left-handed and right-handed composite
spectra, (\ref{45}) and (\ref{45'}), respectively. The similar vectorlike
submultiplets in them get the gravitationally induced Dirac masses (\ref{hm}%
) and (\ref{hm'}), while chiral multiplets receive Majorana masses from the
family symmetry breaking in (\ref{rn}) and (\ref{rn1}). They all decouple at
the scale $M_{GUT}$ where the $SU(5)$ and $SU(3)_{F}$ break, and below this
scale there are only left the three standard families of massless quarks and
leptons. They in turn get their masses at the scale $M_{SM}$ where the
Standard Model breaks down.

And the last interesting scenario could be the case with the $SU(3)_{F}$
family symmetry scale taken as the effective scale in the theory, $\mathcal{M%
}$ $\sim M_{F}$, which could be as low as possible. As we mentioned above,
in order not to come in conflict with possible flavor-changing processes
this scale has to be at least of the order $10^{5}$ $GeV$. At the same time,
to avoid an inadmissible mixing of lepton doublets with those in heavy
vectorlike multiplets (\ref{fr}) masses of the latters have also to be of
the family scale order or less, $M_{V}\lesssim M_{F}$. So, one has for all
scales involved%
\begin{equation}
M_{LR}\sim \Lambda _{MC}\sim M_{Pl}\text{ , \ }M_{GUT}\gtrsim M_{F}\gtrsim
M_{V}\gtrsim 10^{5}GeV  \label{sc3}
\end{equation}%
that for the relatively low family scale $M_{F}$ implies strong hierarchy
between compositeness scale $\Lambda _{MC}$ and an effective theory scale $%
\mathcal{M}$. Meanwhile, the low gravitational conversion scale $M_{V}$ can
be readily reached by only a slight lowering of the metacolor scale in (\ref%
{hm}) down to $\Lambda _{MC}\gtrsim 5\cdot 10^{17}GeV$. Eventually, one has,
apart the three families of ordinary quarks and leptons with masses at the
electroweak scale $M_{SM}$, the heavy chiral multiplets (\ref{fm}) and heavy
vectorlike multiplets (\ref{fr}) with the Majorana and Dirac masses
(respectively) in the interval $(10^{5}-10^{16})$ $GeV$. Indeed, these extra
heavy states might in principle affect the gauge coupling unification. The
point is, however, that they all are presented by the unsplit multiplets of
the $SU(5)\times SU(3)_{F}$ symmetry whose components have the same mass.
Therefore, they may influence unification of the $SU(5)$ gauge coupling
constants in the two-loop approximation only that can be largely ignored in
both ordinary and supersymmetric theories.

Note that none of the above scenarios proposes the further unification of
gauge coupling constants $g_{5}$ and $g_{3F}$ of the $SU(5)$ and $SU(3)_{F}$%
, respectively, at the $SU(8)_{MF}$ unification scale $M_{LR}$. Indeed, as
was mentioned above, the $SU(8)_{MF}$ does not exist for composites due to
anomalies appearing at the large distances. Thus, the $SU(8)_{MF}$ gauge
constant $g_{8}$ is only applicable to gauge interactions of preons and may
substantially differ from gauge coupling constants $g_{5}$ and $g_{3F}$ of
composites. Nonetheless, some significant impact on the $g_{8}$ coupling
constant running may appear from the threshold effects at the compositeness
scale $\Lambda _{MC}$ where the preons themselves come into play. An
appropriate running would make it possible for the $SU(8)_{MF}$ metaflavor
theory of preons to be further unified with gravity at the Planck scale $%
M_{Pl}$. Anyway, a coupling unification in the preon model for quarks and
leptons deserves a special consideration that we plan to address elsewhere.

Meanwhile, one may conclude that, together with an origin of the standard $%
SU(5)$ grand unification, the most important prediction of the left-right
preon model\ considered here is, indeed, an existence of the local family
symmetry $SU(3)_{F}$ for quark-lepton generations. In fact, all quarks and
leptons with the both left-handed and right-handed components appear as
antitriplets of the $SU(3)_{F}$, as is readily seen from their submultiplet $%
(\overline{5}+10,\overline{3})_{L}$ itself in (\ref{fm}). This means that
the $SU(3)_{F}$ is in essence a chiral symmetry, rather than a vectorlike
symmetry such as, for example, the conventional color symmetry $SU(3)_{C}$.
We briefly sketch the $SU(3)_{F}$ family symmetry and some of its basic
applications below.

\section{The family symmetry SU(3)$_{F}$}

The flavor mixing of quarks and leptons is certainly one of the major
problems that presently confronts particle physics. Many attempts have been
made to interpret the pattern of this mixing in terms of various family
symmetries - discrete or continuous, global or local. Among them, the chiral
family symmetry $SU(3)_{F}$ introduced first in the preon model framework
\cite{jp} and then developed by its own by many authors [22-26, 28-31] seems
most promising.

Generically, the chiral family symmetry $SU(3)_{F}$ possesses the following
four distinctive features:

(i) It provides a natural explanation of the number three of observed
quark-lepton families correlated with three species of massless or light ($%
m_{\nu }<M_{Z}/2$) neutrinos contributing to the invisible $Z$ boson partial
decay width;

(ii) Its local nature conforms with the other local symmetries of the
Standard Model, such as the weak isospin symmetry $SU(2)_{W}$ or color
symmetry $SU(3)_{C}$, thus leading to the family-unified SM with a total
symmetry $SM\times SU(3)_{F}$;

(iii) Its chiral nature, according to which both the left-handed and
right-handed quarks and leptons are proposed to be fundamental antitriplets
of the $SU(3)_{F}$, provides the hierarchical mass spectrum of quark-lepton
families as a result of a spontaneous symmetry breaking at some high scale $%
M_{F}$ which could in principle located in the range from $10^{5}$ $GeV$ (to
properly suppress the flavor-changing processes) up to the grand unification
scale $M_{GUT}$ and even higher. Actually, any family symmetry should be
completely broken in order to conform with reality at lower energies.
Meanwhile, this symmetry should be chiral, rather than vectorlike, since a
vectorlike symmetry would not forbid the family invariant masses, thus
leading in general to uniform rather than hierarchical mass spectra.
Interestingly, both known examples of local vectorlike symmetries,
electromagnetic $U(1)_{EM}$ and color $SU(3)_{C}$, appear to be exact
symmetries, while all chiral symmetries including conventional grand
unifications $SU(5)$, $SO(10)$ and $E(6)$ appear broken;

(iv) Thereby, due to its chiral structure, the $SU(3)_{F}$ admits a natural
unification with all known GUTs in a direct product form, both in an
ordinary and supersymmetric framework, thus leading to the family-unified
GUTs, $GUT\times SU(3)_{F}$, beyond the Standard Model.

So, if one takes these natural criteria seriously, any other candidate for
flavor symmetry, except for the local chiral $SU(3)_{F}$ symmetry, can be
excluded. Indeed, the $U(1)$ family symmetry does not satisfy the criterion
(i) and is in fact applicable to any number of quark-lepton families. Also,
the $SU(2)$ family symmetry can contain, besides two light families treated
as its doublets, any number of additional (singlets or new doublets of $%
SU(2) $) families. All global non-Abelian symmetries are excluded by the
criterion (ii), while the vectorlike symmetries are excluded by the last
criteria (iii) and (iv).

Among applications of the $SU(3)_{F}$ symmetry, the most interesting ones
related to description, both in the Standard Model and GUTs, of the quark
and lepton masses and mixings \cite{su3} including the neutrino masses and
oscillations \cite{ber1}. Indeed, spontaneous breaking of this symmetry
gives some guidance to the observed hierarchy between elements of the
quark-lepton mass matrices and presence of texture zeros in them, that in
the preon model framework we schematically discussed above in the section 6.
Remarkably, the $SU(3)_{F}$ invariant Yukawa couplings like those presented
in (\ref{udd}, \ref{ud'}) are always accompanied by an accidental global
chiral $U(1)$ symmetry \cite{q84} which can be identified with the
Peccei-Quinn symmetry \cite{peccei}, thus giving a solution to the strong $%
CP $ problem . For the relatively low family scale $M_{F}$,\ the $SU(3)_{F}$
gauge bosons will also enter into play so that there may become important
many flavor-changing rare processes \cite{jon0} including some of their
astrophysical consequences \cite{m}. In the framework of supersymmetric
theories \cite{q86}, both SM and GUT, the family $SU(3)_{F}$ symmetry hand
in hand with hierarchical masses and mixings for quarks and leptons leads to
an almost uniform mass spectrum for their superpartners with a high degree
of flavor conservation, that makes this symmetry existence even more
significant in the SUSY case. The special sector of applications is related
to a new type of topological defects - flavored cosmic strings and monopoles
appearing due to the spontaneous violation of the $SU(3)_{F}$ which may be
considered as possible candidates for the cold dark matter in the Universe
\cite{def}.

In this context, the question naturally arises whether the $SU(3)_{F}$
family symmetry has its origin in the preon model considered here or it
rather appears as an independently postulated symmetry. This may depend in
fact on the basic scenarios considered in the section 6.5. The most crucial
difference between these two cases is related to an existence in the preon
model of some heavy $SU(5)\times SU(3)_{F}$ multiplets located at scales
from $O(100)$ $TeV$ up to the Planck mass. They are the chiral multiplets (%
\ref{fm}) with the family scale $M_{F}$ order masses and vectorlike
multiplets (\ref{fr}) with the gravitationally induced masses being
presumably of the same order. If they are relatively light, they may be
directly observed and they can also influence the $SU(5)$ gauge coupling
unification. If they are too heavy, they can still strongly affect the mass
matrices for down quarks and leptons that eventually leads to a significant
violation of unitarity in the lepton CKM matrix. Moreover, for the high
scale family symmetry one has in the preon model some actual candidates for
heavy neutrinos that provides, as was shown in the section 6.3, some natural
see-saw mechanism for physical neutrinos.

\section{Conclusion and outlook}

We started by stating that there is still left a serious problem in particle
physics related to classification of all observed quark-lepton families.
This may motivate us to continue seeking a solution in some subparticle or
preon models for quarks and leptons. In this connection, it is worth
underlining at the outset that none of the presently popular $SU(5)$, $%
SO(10) $ and $E(6)$ GUTs satisfies a somewhat evident criterion of possible
elementarity of quarks and leptons. It seems likely that, if they were
elementary they all should be contained in a single fundamental
representation of grand unified symmetry, rather than in a set of its
representations. This, as well as replication of quantum numbers carried by
quarks and leptons, could tell about their composite structure formed by
preons which might be elementary carriers of those quantum numbers. We have
shown above that in accordance with these heuristic arguments the preon
model may under certain natural conditions determine a local metaflavor $%
SU(8)_{MF}$ symmetry as a basic internal symmetry of the physical world at
small distances that is then passed on to large distances in the terms of
the $SU(5)\times SU(3)_{F}$ symmetry for composite quarks and leptons.

Let us now recall all stages we passed to come to the main results presented
here. We started with the $2N$ left-handed and right-handed preons, $P_{iL}^{%
\boldsymbol{a}}$ and\ $Q_{iR}^{\boldsymbol{a}^{\prime }}$ ($i=1,\ldots ,N$; {%
$\boldsymbol{a}$}, $\boldsymbol{a}^{\prime }=1,...,n$), possessing some
local metaflavor symmetry $SU(N)_{MF}$ and chiral metacolor symmetry $%
SO(n)_{MC}^{L}\times SO(n)_{MC}^{R}$ with numbers of metaflavors ($N$) and
metacolors ($n$) not yet determined. We argued that these preons also
possess the chiral symmetry $K(N)$ (\ref{ch}, \ref{ch'}) in the limit when
their gauge $SU(N)_{MF}$ metaflavor interactions are switched off. Though
these interactions break the $K(N)$ symmetry, they are typically too weak at
the metacolor confining distances to influence the formation of composites.
Requiring a preservation of an appropriate chiral symmetry in the bound
state spectrum, provided that this spectrum is only given by some single
representation of $K(N)$ rather than a set of its representations, we found
that just the chiral symmetry $K(8)$ (\ref{881}) is solely selected as a
universal chiral symmetry at all distances for an asymptotically free $%
SO(5)_{MC}^{L}\times SO(5)_{MC}^{R}$ metacolor theory. Simultaneously, this
determines the effective local metaflavor symmetry $SU(8)_{MF}$ that could
be in principle\ observed at large distances through the low-lying
composites emerged. Indeed, any possible extra $N-8$ metaflavors,\ which are
not provided by a preserved chiral symmetry, could only emerge in superheavy
composites with the metacolor scale $\Lambda _{MC}$ order masses, thus
appearing unobserved at a laboratory scale. However, such $L$-$R$ symmetric $%
SU(8)_{MF}$ metaflavor theory certainly is pure vectorlike for the identical
$L$-preon and $R$-preon composite multiplets involved that is inadmissible.
Actually, this means that, while preons are left massless being protected by
their own metacolors, the composites being metacolor singlets will pair up
and acquire heavy Dirac masses. It is rather clear that such a theory is
meaningless unless the $L$-$R$ symmetry is partially broken that seems to be
a crucial point in our model. In this connection, some natural mechanism for
spontaneous $L$-$R$ symmetry breaking caused by the simultaneously emerged
composite scalars has been proposed. According to it, while nothing happens
with the left-handed preon composites still completing the total multiplets
of the $SU(8)_{L}$, the right-handed preon composites will only form some
particular submultiplets of the $[SU(5)\times SU(3)]_{R}$ symmetry. As a
result, the triangle anomalies corresponding to generators of the $SU(8)_{R}$%
, other than those of its $[SU(5)\times SU(3)]_{R}$ subgroup, are left
uncompensated that causes the proper decreasing of the chiral symmetry $K(8)$
shown in (\ref{l-r}). Meanwhile, the $L$-$R$ symmetric metaflavor $SU(8)_{MF}
$ theory of preons breaks down to an effective chiral $SU(5)\times SU(3)_{F}$
theory for composites containing the conventional $SU(5)$ GUT times an extra
local family symmetry $SU(3)_{F}$ describing the three standard quark-lepton
families. Though the tiny radius of compositeness for universal preons
composing both quarks and leptons makes it impossible to immediately confirm
their composite nature, a few special $SU(5)\times SU(3)_{F}$ multiplets of
extra composite fermions, predicted by the theory at the energies from $%
O(100)$ $TeV$ up to the Planck scale, may appear of an actual experimental
interest. Some of them can be directly observed, the others populate the $%
SU(5)$ GUT desert. Due to their mixing with ordinary quark-lepton families
there may emerge a significant violation of unitarity in the lepton CKM
matrix depending on the interplay between the compositeness scale and scale
of the family symmetry $SU(3)_{F}$.

Let us also briefly comment on some ways along which these results can be
extended. The first and immediate one could be a supersymmetric extension of
the present model, though many other supersymmetric preon models have been
proposed \cite{moh, ds}. We introduced above some metaflavorless preons
called sterilons which would be in fact gaugino (metagluino) in the
supersymmetrized metacolor theory. This can be only considered as the first
step which may be then continued totally within a supersymmetric context.
Generally, as a conventional SUSY still remains undiscovered, one could
think that it may be generically related to preons rather than quarks and
leptons in the Standard Model. And, as a possible result, one might only
expect an existence of some effective SUSY at energies being much higher
than the electroweak scale.

Another important question we have not yet considered here is an emergence
of composite vector field in the theory. Generally, the vector composites
have to be heavy with masses of the order of the compositeness scale $%
\Lambda _{MC}$ unless they appear as effective gauge fields. The point is,
however, that some massless composite vector fields could nonetheless appear
in a theory as the vector Goldstone bosons related to spontaneous violation
of Lorentz invariance. Such violation could in principle appear through the
multi-preon interactions similar to those given in the section 5.2.
Actually, one could start with a global (rather than local) metaflavor
symmetry $SU(N)_{MF}$ which is then converted into the local one through the
contact multi-preon interactions \cite{jb} or some nonlinear constraint put
on the preon currents (see in this connection \cite{nam} and the later works
\cite{cfn}). If so, the quarks and leptons, on the one hand, and the
mediator gauge fields (photons, weak bosons, gluons etc.), on the other,
could be composed at the same order distances determined by the preon
confinement scale $\Lambda _{MC}$. In other words, there may be a high
energy limit to the division of matter beyond which one can not go. Indeed,
a conventional division of matter from atoms to quarks is naturally related
to the fact that matter is successively divided, whereas the mediator gauge
fields are left intact. However, situation may be drastically changed if
these spontaneously emerging gauge fields become composite as well. Note
also that closeness of the compositeness scale $\Lambda _{MC}$ to the Planck
scale $M_{Pl}$ in the present model seems to make it in principle possible
for the $SU(8)_{MF}$ metaflavor theory to be further unified with gravity,
particularly in the case when a graviton, like other gauge fields, is
presumably composite. This and other things mentioned above seem to be
especially interesting and worth further pursuing.

\section*{ Acknowledgments}

I am grateful to many people who had significantly contributed to the ideas
presented here during the years when they were developed, especially to A.A.
Anselm, V.N. Gribov, S.G. Matinyan and V.I. Ogievetsky, as well as to my
collaborators Z.G. Berezhiani, O.V. Kancheli, H.B. Nielsen and K.A.
Ter-Martirosyan.

\end{document}